\documentclass[12pt]{article}

\usepackage{amsthm,amscd}
\usepackage{mathrsfs}
\usepackage{verbatim}
\usepackage{amsfonts,amsmath,latexsym,amssymb,txfonts,bm}
\usepackage[american]{babel}

\makeatletter
\renewcommand{\@evenhead}{\raisebox{0pt}[\headheight][0pt]{\vbox{\hbox
to \textwidth{\thepage\hfil\strut\textit{\leftmark}}\hrule}}}
\renewcommand{\@oddhead}{\raisebox{0pt}[\headheight][0pt]{\vbox{\hbox
to \textwidth{\textit{\rightmark}\hfil\strut\thepage}\hrule}}}

\def\timenow{%
\@tempcnta=\time \divide\@tempcnta by 60 \number\@tempcnta:\multiply
\@tempcnta by 60 \@tempcntb=\time \advance\@tempcntb by -\@tempcnta
\ifnum\@tempcntb <10 0\number\@tempcntb\else\number\@tempcntb\fi}

\makeatother

\renewcommand\theequation{\thesection.\arabic{equation}}

\def\II{{\mathbb I}}
\def\RR{{\mathbb R}}

\def\Tr{\mathrm{ Tr\,}}

\def\vol{\mathrm{ vol\,}}

\def\be{\begin{equation}}
\def\ee{\end{equation}}
\def\bea{\begin{eqnarray}}
\def\eea{\end{eqnarray}}
\def\bed{\begin{definition}{\ }}
\def\eed{\end{definition}}
\def\bd{\begin{description}}
\def\ed{\end{description}}
\def\bc{\begin{center}}
\def\ec{\end{center}}

\newtheorem{lemma}{Lemma}

\newtheorem{definition}{Definition}

\def\sideremark#1{\ifvmode\leavevmode\fi\vadjust{\vbox to0pt{\vss
\hbox to 0pt{\hskip\hsize\hskip1em
\vbox{\hsize2cm\tiny\raggedright\pretolerance10000
\noindent #1\hfill}\hss}\vbox to8pt{\vfil}\vss}}}

\begin{document}

\begin{titlepage}
\thispagestyle{empty}
\null

\hspace*{50truemm}{\hrulefill}\par\vskip-4truemm\par
\hspace*{50truemm}{\hrulefill}\par\vskip5mm\par
\hspace*{50truemm}{{\large\sc New Mexico Tech {\rm
(October 27, 2008)
}}}
\vskip4mm\par
\hspace*{50truemm}{\hrulefill}\par\vskip-4truemm\par
\hspace*{50truemm}{\hrulefill}
\par
\bigskip
\bigskip
%\par
%\hspace*{50truemm}{\LARGE\textbf{\textsf{DRAFT}}}
%\par
%\vspace{3cm}
\vfill
\centerline{\huge\bf Non-perturbative Heat Kernel Asymptotics}
\bigskip
\centerline{\huge\bf on Homogeneous Abelian Bundles}
%\bigskip
%\bigskip
\bigskip
\bigskip
\centerline{\Large\bf Ivan G. Avramidi and Guglielmo Fucci}
\bigskip
\centerline{\it New Mexico Institute of Mining and Technology}
\centerline{\it Socorro, NM 87801, USA}
\centerline{\it E-mail: iavramid@nmt.edu, gfucci@nmt.edu}
\bigskip
\medskip
\vfill

{\narrower
\par
% ABSTRACT
We study the heat kernel for a Laplace type partial differential operator
acting on smooth sections of a complex vector bundle with the structure
group $G\times U(1)$
over a Riemannian manifold $M$ without boundary. The total
connection on the vector bundle naturally splits into a $G$-connection and a
$U(1)$-connection, which is assumed to have a parallel curvature $F$. We find
a new local short time asymptotic expansion of the off-diagonal heat kernel
$U(t|x,x')$ close to the diagonal of $M\times M$ assuming the curvature $F$ to
be of order $t^{-1}$. The coefficients of this expansion are polynomial
functions in the Riemann curvature tensor (and the curvature of the
$G$-connection) and its derivatives with universal coefficients depending in a
non-polynomial but analytic way on the curvature $F$, more precisely, on $tF$.
These functions generate all terms quadratic and linear in the Riemann
curvature and of arbitrary order in $F$ in the usual heat kernel coefficients.
In that sense, we effectively sum up the usual short time heat kernel
asymptotic expansion to all orders of the curvature $F$. We compute the first
three coefficients (both diagonal and off-diagonal) of this new asymptotic
expansion.

\par}
%\vfill

{\vbox{\hrule\vspace{3pt}\hfil
{\scriptsize\textit{\hfill\hfill
\jobname.tex; \today; \timenow}}
\hfil}}

\end{titlepage}

%=================================================================

\section{Introduction}
\setcounter{equation}0

The heat kernel is one of the most powerful tools in quantum field theory and
quantum gravity as well as mathematical physics and differential geometry (see
for example
\cite{gilkey95,vassile03,avramidi00,avramidi02,avramidi99,kirsten01,vandeven98,
hurt83} and further references therein). It is of particular importance because
the heat kernel methods give a framework for manifestly covariant calculation
of a wide range of relevant quantities in quantum field theory like one-loop
effective action, Green's functions, effective potential etc.

Unfortunately the exact computation of the heat kernel can be carried out only
for exceptional highly symmetric cases when the spectrum of the operator is
known exactly, (see \cite{camporesi90,hurt83,kirsten01} and the references in
\cite{avramidi96,avramidi08a,avramidi08,avramidi08b}). Although these special
cases are very important, in quantum field theory we need the effective action,
and, therefore, the heat kernel for general background fields. For this reason
various approximation schemes have been developed. One of the oldest methods is
the Minackshisundaram-Pleijel short-time asymptotic expansion of the heat
kernel as $t\to 0$ (see the references in
\cite{gilkey95,avramidi91,vassile03}).

Despite its enormous importance, this method is essentially perturbative. It is
an expansion in powers of the curvatures $R$ and their derivatives and, hence,
is inadequate for large curvatures when $tR\sim 1$. To be able to describe the
situation when at least some of the curvatures are large one needs an
essentially {\it non-perturbative approach}, which effectively sums up in the
short time asymptotic expansion of the heat kernel an infinite series of terms
of certain structure that contain large curvatures (for a detailed analysis see
\cite{avramidi94,avramidi97} and reviews \cite{avramidi99,avramidi02}).
For example, the partial summation of higher derivatives enables one to obtain
a non-local expansion of the heat kernel in powers of curvatures (high-energy
approximation in physical terminology). This is still an essentially
perturbative approach since the curvatures (but not their derivatives) are
assumed to be small and one expands in powers of curvatures.

On another hand to study the situation when curvatures (but not their
derivatives) are large (low energy approximation) one needs an essentially {\it
non-per\-tur\-bative approach}. A promising approach to the calculation of the
low-energy heat kernel expansion was developed in non-Abelian gauge theories
and quantum gravity in
\cite{avramidi93,avramidi94,avramidi94a,avramidi95,avramidi95a,avramidi96,
avramidi08b,avramidi08,avramidi08a}. While the papers
\cite{avramidi93,avramidi94,avramidi95,avramidi95a} dealt with the parallel
$U(1)$-curvature (that is, constant electromagnetic field) in flat space, the
papers \cite{avramidi94a,avramidi96,avramidi08b} dealt with symmetric spaces
(pure gravitational field in absence of an electromagnetic field). The
difficulty of combining the gauge fields and gravity was finally overcome in
the papers \cite{avramidi08,avramidi08a}, where homogeneous bundles
with parallel curvature on symmetric spaces was studied.

In this paper we compute the heat kernel for the covariant Laplacian with a
large parallel $U(1)$ curvature $F$ in a Riemannian manifold (that is, strong
covariantly constant electromagnetic field in an arbitrary gravitational
field). Our aim is to evaluate the first three coefficients of the heat kernel
asymptotic expansion in powers of Riemann curvature $R$ but \emph{in all
orders} of the $U(1)$ curvature $F$. This is equivalent to a partial summation
in the heat kernel asymptotic expansion as $t\to 0$ of all powers of $F$ in
terms which are linear and quadratic in Riemann curvature $R$.

%====================================================
\section{Setup of the Problem}
\setcounter{equation}0

Let $M$ be a $n$-dimensional compact Riemannian manifold without boundary and
${\cal S}$ be a complex vector bundle over $M$ realizing a representation of
the group $G\otimes U(1)$. Let $\varphi$ be a section of the bundle ${\cal S}$
and $\nabla$ be the total connection on the bundle $\mathcal{S}$ (including the
$G$-connection as well as the ${\rm U}(1)$-connection). Then the commutator of
covariant derivatives defines the curvatures \begin{equation}\label{0}
[\nabla_\mu,\nabla_\nu]\varphi=({\cal R}_{\mu\nu}+i F_{\mu\nu})\varphi\;,
\end{equation} where ${\cal R}_{\mu\nu}$ is the curvature of the $G$-connection
and $F_{\mu\nu}$ is the curvature of the $U(1)$-connection (which will be also
called the electromagnetic field).

In the present paper we consider a second-order Laplace type partial
differential operator,
\begin{equation}\label{1}
\mathscr{L}=-\Delta, \qquad \Delta=g^{\mu\nu}\nabla_\mu\nabla_\nu\;.
\end{equation}
The heat kernel for the operator $\mathscr{L}$ is defined as
the solution of the heat equation
\begin{equation}
\left(\partial_{t}+\mathscr{L}\right)U(t|x,x^{\prime})=0\;,
\end{equation}
with the initial condition
\begin{equation}
U(0|x,x^{\prime})=\mathcal{P}(x,x^{\prime})\delta(x,x^{\prime})\;.
\end{equation}
where $\delta(x,x^{\prime})$ is the covariant scalar delta function and
$\mathcal{P}(x,x^{\prime})$ is the operator of parallel transport of the
sections of the bundle ${\cal S}$ along the geodesic from the point $x'$ to the
point $x$.

The spectral properties of the operator $\mathscr{L}$ are described in terms of
the spectral functions, defined in terms of the $L^{2}$ traces of some
functions of the operator $\mathscr{L}$, such as the zeta-function $
\zeta(s)=\Tr\,\mathscr{L}^{-s}, $ and the heat trace
\be
\textrm{Tr}\;\exp(-t\mathscr{L})=
\int\limits_{M}d\vol\;\textrm{tr}\;U^{\rm diag}(t)\,,
\ee
where $d\vol=g^{1/2}dx$ is the Riemannian volume element with $g=\det
g_{\mu\nu}$ and $\textrm{tr}$ denotes the fiber trace.
Here and everywhere below the
diagonal value of any two point quantity $f(x,x')$ denotes the coincidence
limit as $x\to x'$, that is,
\be
f^{\rm diag}=f(x,x)\,.
\ee

It is well known \cite{gilkey95} that the heat kernel has the
asymptotic expansion as $t\rightarrow 0$ (see also
\cite{avramidi91,avramidi99,avramidi00,vassile03})
\begin{equation}
U(t|x,x^{\prime})\sim(4\pi t)^{-n/2}
\mathcal{P}(x,x^{\prime})
\Delta^{1/2}(x,x^{\prime})
\exp\left[-\frac{\sigma(x,x^{\prime})}{2t}\right]
\sum_{k=0}^{\infty}t^{k}a_{k}(x,x^{\prime})\;,
\end{equation}
where $\sigma(x,x^{\prime})$ is the geodesic interval (or the world function)
defined as one half the square of the geodesic distance between the points $x$
and $x^{\prime}$ and $\Delta(x,x^{\prime})$ is the Van Vleck-Morette
determinant. The coefficients $a_{k}(x,x^{\prime})$ are called the off-diagonal
heat kernel coefficients.

The heat kernel diagonal and the heat trace have the
asymptotic expansion as $t\rightarrow 0$ \cite{avramidi91,vassile03}
\be
U^{\rm diag}(t)\sim(4\pi t)^{-n/2}
\sum_{k=0}^{\infty}t^{k}a^{\rm diag}_{k}\;,
\ee
\begin{equation}
\textrm{Tr}\;\exp(-t\mathscr{L})
\sim(4\pi t)^{-n/2}\sum_{k=0}^{\infty}t^{k}A_{k}\;,
\end{equation}
where
\be
a_k^{\rm diag}=a_k(x,x)\,
\ee
and
\begin{equation}
A_{k}=\int\limits_{M}d\vol\;\textrm{tr}\;a^{\rm diag}_{k}\;.
\end{equation}
The coefficients $A_{k}$ are called the global heat kernel coefficients;
they are spectral invariants of the operator $\mathscr{L}$.

The diagonal heat kernel coefficients $a^{\rm diag}_k$ are polynomials in the
jets of the metric, the $G$- connection and the $U(1)$-connection; in other
words, in the curvature tensors and their derivatives. Let us symbolically
denote the jets of the metric and the $G$-connection by
\be
R_{(n)}=\left\{\nabla_{(\mu_1}\cdots\nabla_{\mu_{n}}
R^a{}_{\mu_{n+1}}{}^b{}_{\mu_{n+2})}\,,
\;
\nabla_{(\mu_1}\cdots\nabla_{\mu_{n}}
{\cal R}^a{}_{\mu_{n+1})}
\right\}
\;,
\ee
and the jets of the $U(1)$ connection by
\be
F_{(n)}=\nabla_{(\mu_1}\cdots\nabla_{\mu_{n}}F^a{}_{\mu_{n+1})}\,.
\ee
Here and everywhere below the parenthesis indicate complete symmetrization over
all indices included.

By counting the dimension it is easy to describe the general structure of the
coefficients $a^{\rm diag}_k$. Let us introduce the multi-indices of nonnegative integers
\be
{\bf i}=(i_1,\dots,i_m), \qquad
{\bf j}=(j_1,\dots,j_l)\,.
\ee
Let us also denote
\be
|{\bf i}|=i_1+\cdots+i_m,\qquad
|{\bf j}|=j_1+\cdots+j_l\,.
\ee
Then symbolically
\be
a^{\rm diag}_k=
\sum_{N=1}^k
\sum_{l=0}^N\sum_{m=0}^{N-l}
\;
\sum\limits_{{{\bf i},{\bf j}\ge 0}\atop{|{\bf i}|+|{\bf j}|+2N=2k}}
C_{(k,l,m),{\bf i},{\bf j}} F_{(j_1)}\cdots F_{(j_l)}\;
R_{(i_1)}\cdots R_{(i_m)}\,,
\label{216zza}
\ee
where $C_{(k,l,m),{\bf i},{\bf j}}$ are some universal constants.

The lower order diagonal heat kernel coefficients are well known
\cite{gilkey95,avramidi91,avramidi00}
\bea
a^{\rm diag}_0&=&1\,,
\\[10pt]
a^{\rm diag}_1&=&\frac{1}{6}R\,,
\\[10pt]
a^{\rm diag}_2&=&
{1\over 30}\Delta R
+{1\over 72}R^2
-{1\over 180} R_{\mu\nu}R^{\mu\nu}
+{{1}\over {180}}R_{\alpha\beta\mu\nu}R^{\alpha\beta\mu\nu}
\nonumber\\
&&
+{1\over 12}{\cal R}_{\mu\nu}{\cal R}^{\mu\nu}
+{1\over 6}{\cal R}_{\mu\nu}iF^{\mu\nu}
-{1\over 12}F_{\mu\nu}F^{\mu\nu}
\,.
\label{120iga}
\eea
To avoid confusion we should stress that the normalization of the coefficients
$a_k$ differs from the papers \cite{avramidi91,avramidi99,avramidi00}.

In the present paper we study the case of a {\it parallel $U(1)$ curvature}
(covariantly constant electromagnetic field), i.e.
\begin{equation}
\label{2x}
\nabla_{\mu}F_{\alpha\beta}=0\;.
\end{equation}
That is, all jets $F_{(n)}$ are set to zero except the one of order zero, which
is $F$ itself. In this case eq. (\ref{216zza}) takes the form
\be
a^{\rm diag}_k=
\sum_{N=1}^k
\sum_{l=0}^N\sum_{m=0}^{N-l}
\;
\sum\limits_{{{\bf i}\ge 0}\atop{|{\bf i}|+2N=2k}}
C_{(k,l,m),{\bf i}} F^l\;
R_{(i_1)}\cdots R_{(i_m)}\,,
\ee
where $C_{(k,l,m),{\bf i}}$ are now some (other) numerical coefficients.

Thus, by summing up all powers of $F$ in the asymptotic expansion
of the heat kernel diagonal we obtain a {\it new (non-perturbative)
asymptotic expansion}
\be
U^{\rm diag}(t)\sim(4\pi t)^{-n/2}
\sum_{k=0}^{\infty}t^{k}\tilde a^{\rm diag}_{k}(t)\;,
\label{222zza}
\ee
where the coefficients $\tilde a^{\rm diag}_k(t)$ are polynomials in the jets
$R_{(n)}$
\be
\tilde a^{\rm diag}_k(t)=
\sum_{N=1}^k
\sum_{m=0}^{N}
\;
\sum\limits_{{{\bf i}\ge 0}\atop{|{\bf i}|+2N=2k}}
f^{(k)}_{(m,{\bf i})}(t)\;
R_{(i_1)}\cdots R_{(i_m)}\,,
\label{121iga}
\ee
and  $f^{(k)}_{(m,{\bf i})}(t)$ are some universal dimensionless tensor-valued
analytic functions that depend on $F$ only in the dimensionless combination
$tF$.

For the heat trace we obtain then a new asymptotic expansion of the form
\begin{equation}
\textrm{Tr}\;\exp(-t\mathscr{L})
\sim(4\pi t)^{-n/2}\sum_{k=0}^{\infty}t^{k}
\tilde A_{k}(t)\;,
\end{equation}
where
\begin{equation}
\tilde A_{k}(t)=\int\limits_{M}d\vol\,
\textrm{tr}\, \tilde a^{\rm diag}_{k}(t)\;.
\end{equation}

This expansion can be described more rigorously as follows. We rescale the
$U(1)$-curvature $F$ by
\be
F \mapsto F(t)=t^{-1}\tilde F\,,
\ee
so that $tF(t)=\tilde F$ is independent of $t$. Then the operator
$\mathscr{L}(t)$ becomes dependent on $t$ (in a singular way!). However, the
heat trace still has a nice asymptotic expansion as $t\to 0$
\begin{equation}
\textrm{Tr}\;\exp[-t\mathscr{L}(t)]
\sim(4\pi t)^{-n/2}\sum_{k=0}^{\infty}t^{k}
\tilde A_{k}\;,
\end{equation}
where the coefficients $\tilde A_k$ are expressed in terms of $\tilde F=tF(t)$,
and, therefore, are independent of $t$. Thus, what we are doing is the {\it
asymptotic expansion of the heat trace for a particular case of a singular (as
$t\to 0$) time-dependent operator $\mathscr{L}(t)$}.

Let us stress once again that the eq. (\ref{121iga}) should not be taken
literally; it only represents the general structure of the coefficients $\tilde
a^{\rm diag}_k(t)$. To avoid confusion we list below the general structure of
the low-order coefficients in more detail
\bea
\tilde a^{\rm diag}_0(t)&=&f^{(0)}(t)\,,
\\[10pt]
\tilde a^{\rm diag}_1(t)&=&
f^{(1)}_{(1,1)}{}^{\alpha\beta\mu\nu}(t)R_{\alpha\beta\mu\nu}
+f^{(1)}_{(1,2)}{}^{\mu\nu}(t){\cal R}_{\mu\nu}\,,
\\[10pt]
\tilde a^{\rm diag}_2(t)&=&
f^{(2)}_{(1,1)}{}^{\alpha\beta\mu\nu\sigma\rho}(t)
\nabla_{(\alpha}\nabla_{\beta)} R_{\mu\nu\sigma\rho}
+f^{(2)}_{(1,2)}{}^{\alpha\beta\mu\nu}(t)\nabla_{(\alpha}
\nabla_{\beta)}{\cal R}_{\mu\nu}
\nonumber\\[5pt]
&&
+f^{(2)}_{(2,1)}{}^{\alpha\beta\gamma\delta\mu\nu\sigma\rho}(t)
R_{\alpha\beta\gamma\delta}
R_{\mu\nu\sigma\rho}
+f^{(2)}_{(2,2)}{}^{\alpha\beta\mu\nu}(t)
{\cal R}_{\alpha\beta}
{\cal R}_{\mu\nu}
\nonumber\\[5pt]
&&
+f^{(2)}_{(2,3)}{}^{\alpha\beta\mu\nu\sigma\rho}(t){\cal R}_{\alpha\beta}
R_{\mu\nu\sigma\rho}
\,
%\nonumber\\
\eea
with obvious enumeration of the functions.
It is the {\it universal tensor functions} $f^{(i)}_{(l,m)}(t)$ that are of
prime interest in this paper. Our main goal is to compute the functions
$f^{(i)}_{(l,m)}(t)$ for the coefficients $\tilde a^{\rm diag}_0(t)$, $\tilde
a^{\rm diag}_1(t)$ and
$\tilde a^{\rm diag}_2(t)$.

Of course, for $t=0$ (or $F=0$) the coefficients $\tilde a_k(t)$ are equal to
the usual diagonal heat kernel coefficients
\be
\tilde a^{\rm diag}_k(0)=a_k^{\rm diag}\,.
\ee
Therefore, by using the explicit form of the coefficients $a_k^{\rm diag}$
given by (\ref{120iga}) we obtain the initial values for the functions
$f^{(i)}_{(j,k)}$. Moreover, by analyzing the corresponding terms in the
coefficients $a_3^{\rm diag}$ and $a_4^{\rm diag}$ (which are known,
\cite{gilkey95,avramidi91,vandeven98}), one can obtain partial information
about some lower order Taylor coefficients of the functions
$f^{(i)}_{(j,k)}(t)$:
\bea
f^{(0)}(t)&=&1-\frac{1}{12}t^2F_{\mu\nu}F^{\mu\nu}+O(t^3)\,,
\\[10pt]
f^{(1)}_{(1,1)}{}^{\alpha\beta}{}_{\mu\nu}(t)
&=&
\frac{1}{6}\delta^{\alpha}_{[\mu}\delta^\beta_{\nu]}
+O(t)\,,
\\[10pt]
f^{(1)}_{(1,2)}{}^{\mu\nu}(t)&=&
\frac{1}{6}tiF^{\mu\nu}+O(t^2)\,,
%\\[10pt]
\eea
\bea
f^{(2)}_{(1,1)}{}^{\alpha\beta\mu\nu}{}_{\sigma\rho}(t)
&=&
\frac{1}{30}g^{\alpha\beta}\delta^{\mu}_{[\sigma}\delta^\nu_{\rho]}
+O(t)\,,
\\[10pt]
f^{(2)}_{(1,2)}{}^{\alpha\beta}{}_{\mu\nu}(t)&=&
-\frac{1}{15}tiF^{(\alpha}{}_{[\nu}\delta^{\beta)}_{\mu]}
+O(t^2)\,,
\\[10pt]
f^{(2)}_{(2,1)}{}_{\alpha\beta}{}^{\gamma\delta}{}_{\mu\nu}{}^{\sigma\rho}(t)
&=&
\frac{1}{180}g_{\mu[\alpha}g_{\beta]\nu}g^{\sigma[\gamma}g^{\delta]\rho}
-\frac{1}{180}\delta^{[\gamma}_{[\alpha}g_{\beta][\nu}g^{\delta][\rho}
\delta^{\sigma]}_{\mu]}
\nonumber\\
&&+\frac{1}{72}\delta^{\gamma}_{[\alpha}\delta^\delta_{\beta]}
\delta^{\sigma}_{[\mu}\delta^\rho_{\nu]}
+O(t)
\,,
\nonumber\\
&&\\
%[10pt]
f^{(2)}_{(2,2)}{}^{\alpha\beta}{}_{\mu\nu}(t)
&=&
\frac{1}{12}\delta^{\alpha}_{[\mu}\delta^\beta_{\nu]}
+O(t)\,,
\\[10pt]
f^{(2)}_{(2,3)}{}^{\alpha\beta\mu\nu}{}_{\sigma\rho}(0)&=&
-\frac{1}{36}tiF^{\alpha\beta}\delta^{\mu}_{[\sigma}\delta^\nu_{\rho]}
-\frac{1}{30}tiF^{\mu\nu}\delta^{\alpha}_{[\sigma}\delta^\beta_{\rho]}
+\frac{1}{9}\delta^{[\mu}_{[\sigma}tiF^{\nu][\alpha}\delta^{\beta]}_{\rho]}
+O(t^2)
\nonumber\,.
\\[10pt]
\eea
This information can be used to check our final results.

Notice that the global coefficients $\tilde A_k(t)$ have exactly the same form
as the local ones; the only difference is that the terms with the derivatives
of the Riemann curvature do not contribute to the integrated coefficients since
they can be eliminated by integrating by parts and taking into account that $F$
is covariantly constant.

Moreover, we study even more general non-perturbative asymptotic expansion for
the {\it off-diagonal} heat kernel and compute the coefficients of zero, first
and second order in the Riemann curvature. We will show that there is a
{\it new non-perturbative asymptotic expansion} of the off-diagonal heat kernel
as $t\to 0$ (and $F=t^{-1}\tilde F$, so that $tF$ is fixed) of the form
\be
U(t|x,x')\sim {\cal P}(x,x')\Delta^{1/2}(x,x')U_0(t|x,x')
\sum_{k=0}^\infty t^{k/2} b_k(t|x,x')
\ee
where $U_0$ is an analytic function of $F$ such that for $F=0$
\be
U_0(t|x,x')\Bigg|_{F=0}=(4\pi t)^{-n/2}
\exp\left[-\frac{\sigma(x,x')}{2t}\right]\,.
\ee
Here $b_k(t|x,x')$ are analytic functions of $t$ that depend on $F$ only in the
dimensionless combination $tF$. Of course, for $t=0$ they are equal to the
usual heat kernel coefficients, that is,
\be
b_{2k}(0|x,x')=a_k(x,x')\,,
\qquad
b_{2k+1}(0|x,x')=0.
\ee
Moreover, we will show below that the odd-order coefficients vanish
not only for $t=0$ and any $x\ne x'$ but also for any $t$ and $x=x'$,
that is, on the diagonal,
\be
b_{2k+1}^{\rm diag}(t)=0\,.
\ee
Thus, the heat kernel diagonal has the asymptotic expansion
(\ref{222zza}) as $t\to 0$ with
\be
\tilde a^{\rm diag}_k(t)=(4\pi t)^{n/2}U_0^{\rm diag}(t)b_{2k}^{\rm diag}(t)\,.
\ee

%=============================================================

\section{Geometric Framework}
\setcounter{equation}0

Our goal is to study the heat kernel $U(t|x,x')$ in the neighborhood of the
diagonal as $x\to x'$. Therefore, we will expand all relevant quantities in
covariant Taylor series near the diagonal following the methods developed in
\cite{avramidi91,avramidi00,avramidi99,avramidi02}. We fix a point, say
$x^{\prime}$, on the manifold $M$ and consider a sufficiently small
neighborhood of $x^{\prime}$, say a geodesic ball with a radius smaller than
the injectivity radius of the manifold. Then, there exists a unique geodesic
that connects every point $x$ to the point $x^{\prime}$. In order to avoid a
cumbersome notation, we will denote by \emph{Latin letters} tensor indices
associated to the point $x$ and by \emph{Greek letters} tensor indices
associated to the point $x^{\prime}$. Of course, the indices associated with
the point $x$ (resp. $x^{\prime}$) are raised and lowered with the metric at
$x$ (resp. $x^{\prime}$). Also, we will denote by $\nabla_a$ (resp.
$\nabla'_\mu$) covariant derivative with respect to $x$ (resp. $x'$).

We remind below the definition of some of the two-point functions that we will
need in our analysis. First of all, the world function $\sigma(x,x^{\prime})$
is defined as one half of the square of the length of the geodesic between the
points $x$ and $x^{\prime}$. It satisfies the equation
\be
\sigma=\frac{1}{2}u^a u_a=u_\mu u^\mu\,,
\ee
where
\begin{equation}
\label{4}
u_{a}=\nabla_{a}\sigma,
\qquad
u_{\mu}=\nabla^{\prime}_{\mu}\sigma\;.
\end{equation}
The variables $u^\mu$ are nothing but the normal coordinates at the point $x'$.

The Van Vleck-Morette determinant is defined by
\begin{equation}
\label{e}
\Delta(x,x^{\prime})
=g^{-\frac{1}{2}}(x)\det[-\nabla_{a}\nabla^{\prime}_{\nu}
\sigma(x,x^{\prime})]g^{-\frac{1}{2}}(x^{\prime})\;.
\end{equation}
This quantity should not be confused with the Laplacian
$\Delta=g^{\mu\nu}\nabla_\mu\nabla_\nu$. Usually, the meaning of $\Delta$ will
be clear from the context. We find it convenient to parameterize it by
\begin{equation}\label{6b}
\Delta(x,x^{\prime})=\exp[{2\zeta(x,x')}]\;.
\end{equation}
Next, we define the tensor
\begin{equation}
\label{c}
\eta^{\mu}{}_{b}=\nabla_{b}\nabla^{\prime\mu}\sigma\;,
\end{equation}
and the tensor $\gamma^a{}_\mu$ inverse to $\eta^\mu{}_a$ by
\be
\gamma^a{}_\mu\eta^\mu{}_b=\delta^a_b\,,\qquad
\eta^\mu{}_b\gamma^b{}_\nu=\delta^\mu_\nu\,.
\ee

This enables us to define
new derivative operators
by
\begin{equation}
\bar{\nabla}_{\mu}=\gamma^{a}{}_{\mu}\nabla_{a}\;.
\end{equation}
These operators commute when acting on objects that have been
parallel transported to the point $x^{\prime}$
(in other words the objects that do not have Latin indices).
In fact, when acting on such objects these operators are just partial
derivatives with respect to normal coordinate $u$
\be
\bar\nabla_\mu=\frac{\partial}{\partial u^\mu}\,.
\ee
We also define the operators
\begin{equation}\label{10aa}
\mathcal{D}_{\mu}=\bar{\nabla}_{\mu}
-\frac{1}{2}i F_{\mu\alpha}u^{\alpha}\;.
\end{equation}
Obviously, they form the algebra
\begin{equation}\label{16a}
\left[\mathcal{D}_{\mu},\mathcal{D}_{\nu}\right]=iF_{\mu\nu}\;,\qquad \left[\mathcal{D}_{\mu},u^{\nu}\right]=\delta_{\mu}{}^{\nu}\;.
\end{equation}

Next, the parallel displacement operator $\mathcal{P}(x,x^{\prime})$ of sections
of the vector bundle ${\cal S}$ along the geodesic from the point
$x^{\prime}$ to the point $x$ is defined as the solution of the
equation
\begin{equation}
u^{a}\nabla_{a}\mathcal{P}(x,x^{\prime})=0\;,
\end{equation}
with the initial condition
\begin{equation}\label{6c}
\mathcal{P}(x,x)=\mathbb{I}\;,
\end{equation}
where $\II$ is the identity endomorphism of the bundle ${\cal S}$.
Finally, we define the two-point quantity
\begin{equation}\label{10}
\mathscr{A}_{\mu}=\mathcal{P}^{-1}\bar{\nabla}_{\mu}\mathcal{P}\;.
\end{equation}

We remind, here, that
we consider the case of a covariantly constant electromagnetic field, i.e.
\begin{equation}
\label{2xx}
\nabla_{\mu}F_{\alpha\beta}=0\;.
\end{equation}
In this case we find it useful to decompose the quantity $\mathscr{A}_{\mu}$
as
\begin{equation}\label{9a}
\mathscr{A}_{\mu}=-{1\over 2}i F_{\mu\alpha}u^\alpha+\bar{\mathscr{A}}_{\mu}\;.
\end{equation}

By using this machinery we can rewrite the heat kernel as follows.
First of all, the heat kernel can be
presented in the form
\begin{equation}\label{5}
U(t|x,x^{\prime})=\exp\left({-t\mathscr{L}}\right)\mathcal{P}(x,x^{\prime})
\delta(x,x^{\prime})\;,
\end{equation}
which can also be written as
\begin{equation}\label{5a}
U(t|x,x^{\prime})
=\mathcal{P}(x,x^{\prime})\Delta^{\frac{1}{2}}(x,x^{\prime})
\exp(-t\tilde{\mathscr{L}})\delta(u)\;,
\end{equation}
where $\delta(u)$ is the usual delta-function in the normal coordinates
$u^\mu$ (recall that $u^\mu$ depends on $x$ and $x'$ and $u=0$ when $x=x'$)
and
$\tilde{\mathscr{L}}$ is an operator defined by
\begin{equation}
\tilde{\mathscr{L}}=
\mathcal{P}^{-1}(x,x^{\prime})\Delta^{-\frac{1}{2}}(x,x^{\prime})\mathscr{L}
\Delta^{\frac{1}{2}}(x,x^{\prime})\mathcal{P}(x,x^{\prime})\;.
\end{equation}
As is shown in \cite{avramidi91,avramidi00}
the operator $\tilde{\mathscr{L}}$ can be written in the form
\begin{eqnarray}\label{10a}
\tilde{\mathscr{L}}&=&
-\Delta^{\frac{1}{2}}(\bar{\nabla}_{\mu}
+\mathscr{A}_{\mu})
\Delta^{-1}X^{\mu\nu}(\bar{\nabla}_{\nu}
+\mathscr{A}_{\nu})\Delta^{\frac{1}{2}}
\nonumber\\
&=&-\left(\bar{\nabla}_{\mu}
+\mathscr{A}_{\mu}
-\zeta_{\mu}\right)X^{\mu\nu}\left(\bar{\nabla}_{\nu}
+\mathscr{A}_{\nu}
+\zeta_{\nu}\right)
\nonumber\\
&=&-\left({\cal D}_{\mu}
+\bar{\mathscr{A}}_{\mu}-\zeta_{\mu}\right)X^{\mu\nu}
\left({\cal D}_{\nu}
+\bar{\mathscr{A}}_{\nu}+\zeta_{\nu}\right)
\;,
\end{eqnarray}
where
$
\zeta_{\mu}
=\bar{\nabla}_{\mu}\zeta\;.
$

Now, by using these equations and by recalling the formula in (\ref{9a}),
one can rewrite the operator in (\ref{10a}) in another way as follows
\begin{equation}
\tilde{\mathscr{L}}
=-\left(X^{\mu\nu}\mathcal{D}_{\mu}\mathcal{D}_{\nu}
+Y^{\mu}\mathcal{D}_{\mu}+Z\right)\;,
\end{equation}
where
\begin{eqnarray}
\label{6a}
X^{\mu\nu}&=&\eta^{\mu}{}_{a}\eta^{\nu a}\;,
\\[10pt]
Y^{\mu}&=&
(\bar{\nabla}_{\mu}X^{\mu\nu})
+2X^{\mu\nu}\bar{\mathscr{A}}_{\mu}\;,
%\nonumber
\\[10pt]
Z&=&
\bar{\mathscr{A}}_{\mu}X^{\mu\nu}\bar{\mathscr{A}}_{\nu}
-\zeta_{\mu}X^{\mu\nu}\zeta_{\nu}
+(\bar{\nabla}_{\mu}X^{\mu\nu})\bar{\mathscr{A}}_{\nu}
+(\bar{\nabla}_{\mu}X^{\mu\nu})\zeta_{\nu}
\nonumber\\
&+&X^{\mu\nu}\bar{\nabla}_{\mu}\bar{\mathscr{A}}_{\nu}
+X^{\mu\nu}\bar{\nabla}_{\mu}\zeta_{\nu}\;.
\end{eqnarray}

%============================================================
%========================================================================
\section{Perturbation Theory}
\setcounter{equation}0

Our goal is now to develop the perturbation theory for the heat kernel. We need
to identify a small expansion parameter $\varepsilon$ in which the perturbation
theory will be organized as $\varepsilon\to 0$. First of all, we assume that
$t$ is small, more precisely, we require $t\sim\varepsilon^2$. Also, since we
will work close to the diagonal, that is, $x$ is close to $x'$, we require that
$u^\mu\sim \varepsilon$. This will also mean that $\bar\nabla\sim
\varepsilon^{-1}$ and $\partial_t\sim\varepsilon^{-2}$. Finally, we assume that
$F$ is large, that is, of order $F\sim \varepsilon^{-2}$. To summarize,
\be
t\sim \varepsilon^2,\qquad
u^\mu\sim \varepsilon,\qquad
F\sim \varepsilon^{-2}\,.
\ee

%===================================================
\subsection{Covariant Taylor Expansion}

The Taylor expansions of the quantities introduced above have
the form (up to the fifth order) \cite{avramidi91,avramidi00}
\begin{eqnarray}\label{7}
X^{\mu\nu}&=&
g^{\mu\nu}
+\frac{1}{3}R^{\mu}{}_{\alpha}{}^{\nu}{}_{\beta}u^{\alpha}u^{\beta}
-\frac{1}{6}\nabla_{\alpha}
R^{\mu}{}_{\beta}{}^{\nu}{}_{\gamma}u^{\alpha}u^{\beta}u^{\gamma}
+\frac{1}{20}\nabla_{\alpha}\nabla_{\beta}
R^{\mu}{}_{\gamma}{}^{\nu}{}_{\delta}
u^{\alpha}u^{\beta}u^{\gamma}u^{\delta}
\nonumber\\
&+&
{1\over 15}
R^{\mu}{}_{\alpha\lambda\beta}
R^\lambda{}_{\gamma}{}^{\nu}{}_{\delta}u^{\alpha}u^{\beta}u^{\gamma}u^{\delta}
+O(u^{5})\;,
\end{eqnarray}
\begin{eqnarray}\label{8}
\zeta&=&
\frac{1}{12}R_{\alpha\beta}u^{\alpha}u^{\beta}
-\frac{1}{24}\nabla_{\alpha}R_{\beta\gamma}u^{\alpha}u^{\beta}u^{\gamma}
+\frac{1}{80}\nabla_{\alpha}\nabla_{\beta}
R_{\gamma\delta}u^{\alpha}u^{\beta}u^{\gamma}u^{\delta}
\nonumber\\
&+&
\frac{1}{360}R_{\mu\alpha\nu\beta}
R^{\mu}{}_{\gamma}{}^{\nu}{}_{\delta}u^{\alpha}u^{\beta}u^{\gamma}u^{\delta}
+O(u^{5})\;,
\end{eqnarray}
\begin{eqnarray}\label{8a}
\Delta^{\frac{1}{2}}&=&1
+\frac{1}{12}R_{\alpha\beta}u^{\alpha}u^{\beta}
-\frac{1}{24}\nabla_{\alpha}R_{\beta\gamma}u^{\alpha}u^{\beta}u^{\gamma}
+\frac{1}{80}\nabla_{\alpha}\nabla_{\beta}
R_{\gamma\delta}u^{\alpha}u^{\beta}u^{\gamma}u^{\delta}
\nonumber\\
&+&
\frac{1}{288}R_{\alpha\beta}R_{\gamma\delta}u^{\alpha}u^{\beta}u^{\gamma}u^{\delta}
+\frac{1}{360}R_{\mu\alpha\nu\beta}
R^{\mu}{}_{\gamma}{}^{\nu}{}_{\delta}u^{\alpha}u^{\beta}u^{\gamma}u^{\delta}
+O(u^{5})\;.
\end{eqnarray}
\begin{eqnarray}\label{9}
\bar{\mathscr{A}}_{\mu}
&=&
-{1\over 2}\mathcal{R}_{\mu\alpha}u^\alpha
+{1\over 24}R_{\mu\alpha\nu\beta}i F^\nu{}_{\gamma}u^\alpha u^\beta u^\gamma
+\frac{1}{3}\nabla_{\alpha}\mathcal{R}_{\mu\beta}u^{\alpha}u^{\beta}
\nonumber\\
&+&
{1\over 24}R_{\mu\alpha\nu\beta}
\mathcal{R}^\nu{}_\gamma u^\alpha u^\beta u^\gamma
-\frac{1}{8}\nabla_{\alpha}\nabla_{\beta}
\mathcal{R}_{\mu\gamma}u^{\alpha}u^{\beta}u^{\gamma}
\nonumber\\
&-&{1\over 720}R_{\mu\alpha\nu\beta}R^\nu{}_{\gamma\lambda\delta}
i F^\lambda{}_\epsilon u^\alpha u^\beta u^\gamma u^\delta u^\epsilon
+O(u^6)\;.
\end{eqnarray}
We would like to stress that all coefficients of such expansions are evaluated
at the point $x^{\prime}$. Also note that, the expansion for
$\bar{\mathscr{A}_{\mu}}$ is valid in the case of a covariantly constant
electromagnetic field.

%===============================================================
\subsection{Perturbation Theory for the Operator $\mathscr{L}$}

Now, we expand the operator $\tilde{\mathscr{L}}$ in a formal power series in
$\varepsilon$ (recall that ${\cal D}\sim \varepsilon^{-1}$ and $u\sim
\varepsilon$) to obtain
\begin{equation}\label{14}
\mathscr{L}
\sim-\sum_{k=0}^\infty \mathscr{L}_{k}
\;,
\end{equation}
where $\mathscr{L}_k$ are operators of order $\varepsilon^{k-2}$.
In particular,
\bea
\label{15}
\mathscr{L}_0
&=&
{\cal D}^2\;,
\\
\mathscr{L}_1&=&0\,,
\\
\label{16}
\mathscr{L}_{k}
&=&
X_{k}^{\mu\nu}\mathcal{D}_{\mu}\mathcal{D}_{\nu}
+Y_{k}^{\mu}\mathcal{D}_{\mu}
+Z_{k}\;,
\qquad
k\ge 2\,.
\eea
where
\be
{\cal D}^2=g^{\mu\nu}\mathcal{D}_{\mu}\mathcal{D}_{\nu}\;,
\ee
and $X^{\mu\nu}_k$, $Y^\mu_k$ and $Z_k$ are some tensor-valued polynomials in
normal coordinates $u^\mu$.

Note that $X^{\mu\nu}_k$ are homogeneous
polynomials in normal coordinates $u^\mu$ and $F$ of order $\varepsilon^{k}$.
Similarly, $Y^\mu_k\sim \varepsilon^{k-1}$ and $Z_k\sim \varepsilon^{k-2}$. Of
course, here the terms $Fuu$ are counted as of order zero. That is,
they have the form
\bea
X_k^{\mu\nu}
&=&
P_{(1),\;k}^{\mu\nu}(u)\,,
\\
Y_k^\mu
&=&
P_{(2),\;k-1}^\mu
+F_{\alpha\beta}P_{(3),\;k+1}^{\mu\alpha\beta}(u)\,,
\\
Z_k
&=&
P_{(4),\;k-2}
+F_{\alpha\beta}P_{(5),\;k}^{\alpha\beta}(u)
+F_{\alpha\beta}F_{\rho\sigma}P^{\alpha\beta\rho\sigma}_{(6),\;k+2}(u)\,,
\eea
where $P_{(j),\;k}(u)$ are homogeneous tensor valued
polynomials of degree $k$.

By using the covariant Taylor expansions in (\ref{7}), (\ref{9}) and (\ref{8})
we find the explicit expression of the coefficients
\begin{eqnarray}
\label{17}
X^{\mu\nu}_2
&=&
C^{\mu\nu}_{2}{}_{\alpha\beta} u^\alpha u^\beta\;,
\\
Y^\mu_2
&=&
E^{\mu}_{2}{}_{\alpha}u^\alpha
+G^{\mu}_{2}{}_{\alpha\beta\gamma}u^\alpha u^\beta u^\gamma\;,
\\
Z_2
&=&
H_{2\;\alpha\beta}u^\alpha u^\beta+L_{2}\;,
\label{17a}\\[8pt]
%\end{eqnarray}
%\begin{eqnarray}
X^{\mu\nu}_3&=&
C^{\mu\nu}_{3}{}_{\alpha\beta\gamma} u^\alpha u^\beta u^{\gamma}\;,
\\
Y^\mu_3&=&
E^{\mu}_{3}{}_{\alpha\beta}u^\alpha u^{\beta}\;,
\\
Z_3&=&
H_{3\;\alpha}u^\alpha\;,
\label{17aa}
\\[8pt]
%\end{eqnarray}
%\begin{eqnarray}
X^{\mu\nu}_4&=&
C^{\mu\nu}_{4}{}_{\alpha\beta\gamma\delta}
u^\alpha
u^\beta u^\gamma u^\delta\;,
\label{18d}
\\
Y^\mu_4&=&
E^{\mu}_{4}{}_{\;\alpha\beta\gamma}u^\alpha u^\beta u^\gamma
+G^{\mu}_{4}{}_{\;\alpha\beta\gamma\delta\epsilon}
u^\alpha u^\beta u^\gamma u^\delta u^\epsilon\;,
\\
Z_4&=&
H_{4\;\alpha\beta}u^\alpha u^\beta
+L_{4\;\alpha\beta\gamma\delta} u^\alpha u^\beta u^\gamma u^\delta
+O_{4\;\alpha\beta\gamma\delta\epsilon\kappa}
u^\alpha u^\beta u^\gamma u^\delta u^\epsilon u^\kappa\;,
\label{18a}
\end{eqnarray}
where
\begin{eqnarray}
\label{18b}
C^{\mu\nu}_{2}{}_{\alpha\beta}
&=&
{1\over 3}R^{\mu}{}_{(\alpha}{}^{\nu}{}_{\beta)}\;,
\nonumber\\
E^{\mu}_{2}{}_{\alpha}
&=&
-{1\over 3}R^{\mu}{}_{\alpha}
-{\cal R}^\mu{}_\alpha\;,
\nonumber\\
G^{\mu}_{2}{}_{\alpha\beta\gamma}
&=&
-{1\over 12}R^\mu{}_{(\alpha}{}^\nu{}_{\beta} i F_{\gamma)\nu}\;,
\nonumber\\
H_{2\;\alpha\beta}
&=&
-{1\over 24}R_{\mu(\alpha} i F^{\mu}{}_{\beta)}\;,
\nonumber\\
L_{2}
&=&
\frac{1}{6}R\;,
\end{eqnarray}
\begin{eqnarray}\label{18c}
C^{\mu\nu}_{3}{}_{\alpha\beta\gamma}
&=&
-{1\over 6}\nabla_{(\alpha}R^{\mu}{}_{\beta}{}^{\nu}{}_{\gamma)}\;,
\nonumber\\
E^{\mu}_{3}{}_{\alpha\beta}
&=&
\frac{1}{3}\nabla_{(\alpha}R^{\mu}{}_{\beta)}
-\frac{1}{6}\nabla^{\mu}R_{\alpha\beta}
+\frac{2}{3}\nabla_{(\alpha}\mathcal{R}^{\mu}{}_{\beta)}\;,
\nonumber\\
H_{3\;\alpha}
&=&
\frac{1}{3}\nabla_{\mu}\mathcal{R}^{\mu}{}_{\alpha}
-\frac{1}{6}\nabla_{\alpha}R\;,
\end{eqnarray}
\begin{eqnarray}
C^{\mu\nu}_{4}{}_{\alpha\beta\gamma\delta}
&=&
{1\over 15}R^{\mu}{}_{(\alpha|\lambda|\beta}
R^\lambda{}_{\gamma}{}^{\nu}{}_{\delta)}
+\frac{1}{20}\nabla_{(\alpha}\nabla_{\beta}
R^{\mu}{}_{\gamma}{}^{\nu}{}_{\delta)}\;,
\nonumber\\
E^{\mu}_{4}{}_{\;\alpha\beta\gamma}
&=&
-{1\over 15}R^\mu{}_{\nu(\alpha|\lambda|}R^\nu{}_\beta{}^{\lambda}{}_{\gamma)}
-{1\over 60} R^\mu{}_{(\alpha}{}^\nu{}_\beta R_{\gamma)\nu}
-{1\over 4}R^{\mu}{}_{(\alpha}{}^\nu{}_\beta {\cal R}_{|\nu|\gamma)}
\nonumber\\
&+&
\frac{1}{10}\nabla_{(\alpha}\nabla^{\mu}R_{\beta\gamma)}
-\frac{3}{20}\nabla_{(\alpha}\nabla_{\beta}R^{\mu}{}_{\gamma)}
-\frac{1}{4}\nabla_{(\alpha}\nabla_{\beta}\mathcal{R}^{\mu}{}_{\gamma)}\;,
\nonumber\\
G^{\mu}_{4}{}_{\;\alpha\beta\gamma\delta\epsilon}
&=&
{1\over 40}R^\mu{}_{(\alpha|\nu|\beta}R^{\nu}{}_{\gamma}{}^{\lambda}{}_{\delta} i F_{|\lambda|\epsilon)}\;,
\nonumber\\
H_{4\;\alpha\beta}
&=&
{1\over 4}{\cal R}_{\mu(\alpha}{\cal R}^{\mu}{}_{\beta)}
-\frac{1}{30}R_{\mu\alpha}R^{\mu}{}_{\beta}
-\frac{1}{4}\nabla_{(\alpha}\nabla_{|\mu|}\mathcal{R}^{\mu}{}_{\beta)}
+\frac{1}{60}R_{\mu\nu}R^{\mu}{}_{\alpha}{}^{\nu}{}_{\beta}
\nonumber\\
&+&
\frac{1}{60}R_{\mu\lambda\gamma\alpha}R^{\mu\lambda\gamma}{}_{\beta}
+\frac{1}{40}\Delta R_{\alpha\beta}
+\frac{3}{40}\nabla_{\alpha}\nabla_{\beta}R\;,
\nonumber\\
L_{4\;\alpha\beta\gamma\delta}
&=&
-{1\over 80}R_{\mu(\alpha}{}^{\nu}{}_{\beta}
R^{\mu}{}_{\gamma}i F_{|\nu|\delta)}
-{1\over 80}R_{\mu(\alpha|\lambda|\beta}
R^{\lambda}{}_{\gamma}{}^{\mu\nu} i F_{|\nu|\delta)}
-{1\over 24}{\cal R}_{\mu(\alpha}R^\mu{}_\beta{}^\nu{}_\gamma i F_{|\nu|\delta)}\;,
\nonumber\\
O_{4\;\alpha\beta\gamma\delta\epsilon\kappa}
&=&
{1\over 576}R_{\mu(\alpha}{}^{\nu}{}_{\beta} R^{\mu}{}_{\gamma}{}^{\lambda}{}_{\delta}
i F_{|\nu|\epsilon}i F_{|\lambda|\kappa)}\;.
\end{eqnarray}
Here and everywhere below the parenthesis denote the complete symmetrization
over all indices enclosed; the vertical lines indicate the indices excluded
from the symmetrization.

%===========================================================

\subsection{Perturbation Theory for the Heat Semigroup}

Now, by using the perturbative expansion (\ref{14}) of the operator
$\tilde{\mathscr{L}}$ and recalling that $\mathcal{D}^{2}\sim \varepsilon^{-2}$
and $t\sim \varepsilon^2$, we see that the operator $t\mathcal{D}^{2}$ is of
zero order and the operator $t\mathscr{L}_k$, $k\ge 2$, is of (higher) order
$\varepsilon^{k}$. Therefore, we can consider the terms $t{\mathscr{L}}_k$ with
$k\ge 2$ as a perturbation.

In order to evaluate the heat semigroup we utilize
the Volterra series for the exponent of two non-commuting operators.
Let $X$ be an operator and $Y$ be a perturbation (say of order one
in a small parameter). Then
\be
\exp(X+Y)=T\exp X\,,
\ee
where
\be
T=I
+\sum_{k=1}^{\infty}\int\limits_{0}^{1}d\tau_{k}
\int\limits_{0}^{\tau_{k}}d\tau_{k-1}\cdots
\int\limits_{0}^{\tau_{2}}d\tau_{1}\;
\tilde Y(\tau_1)\tilde Y(\tau_2)
\cdots
\tilde Y(\tau_k)
\ee
and
\be
\tilde Y(\tau)=e^{\tau X}Ye^{-\tau X}\;.
\ee

By using the above series for the operator in (\ref{14})
we obtain
\begin{equation}
\label{27a}
\exp(-t\tilde{\mathscr{L}})
=T(t)\;\exp(t\mathcal{D}^{2})\;,
\end{equation}
where $T(t)$ is an operator defined by a formal perturbative
expansion
\begin{eqnarray}
\label{27b}
T(t)\sim
\sum_{k=0}^\infty T_k(t)
\;,
\end{eqnarray}
with $T_k(t)$ being of order $\varepsilon^k$.
Explicitly, up to terms of fifth order we obtain
\bea
T_0(t)&=&I\,,
\\[10pt]
T_1(t)&=&0\,,
\\
\label{27c}
T_2(t)&=&
t\int\limits_{0}^{1}d\tau_{1}\;
V_2(t\tau_1)\,,
\\
%\ee
%\be
T_3(t)&=&
t\int\limits_{0}^{1}d\tau_{1}\;
V_3(t\tau_1)
%\ee
%\bea
\\
T_4(t)&=&
t\int\limits_{0}^{1}d\tau_{1}\;
V_4(t\tau_1)
%\;,
%\label{27c}
+t^2\int\limits_{0}^{1}d\tau_{2}
\int\limits_{0}^{\tau_{2}}d\tau_{1}\;
V_2(t\tau_1)V_2(t\tau_2)
\;,
\label{27d}
\eea
and
\be
V_k(s)=e^{s\mathcal{D}^{2}}
\mathscr{L}_{k}
e^{-s\mathcal{D}^{2}}\,.
\label{331iga}
\ee

%============================================
\subsection{Perturbation Theory for the Heat Kernel}

As we already mentioned above the heat kernel can be computed from the heat
semigroup by using the equation (\ref{5a}). By using the heat semigroup
expansion from the previous section we now obtain the heat kernel in the form
\bea
\label{25a}
U(t|x,x^{\prime})
&\sim &
\mathcal{P}(x,x^{\prime})\Delta^{1/2}(x,x')U_{0}(t|x,x^{\prime})
\sum_{k=0}^\infty t^{k/2}b_k(t|x,x')
\label{54}
\eea
where
\begin{equation}
\label{23}
U_{0}(t|x,x^{\prime})=\exp(t\mathcal{D}^{2})
\delta(u)\;,
\end{equation}
and
\begin{equation}
\label{54az}
b_{k}(t|x,x^{\prime})=t^{-k/2}
U^{-1}_0(t|x,x')T_{k}(t)U_{0}(t|x,x^{\prime})\;.
\end{equation}
Thus, the calculation of the heat kernel coefficients reduces to the
evaluation of the zero-order heat kernel $U_0(t|x,x')$ and to the action of the
differential operators $T_k(t)$ on it.

The zero order heat kernel $U_{0}(t|x,x^{\prime})$ can be evaluated by using
the algebraic method developed in \cite{avramidi93,avramidi94}. First, the heat
semigroup $\exp(t{\cal D}^2)$ can be represented as an average over the
(nilpotent) Lie group (\ref{16a}) with a Gaussian measure
\begin{eqnarray}
\label{22}
\exp(t\mathcal{D}^{2})
=(4\pi t)^{-n/2}J(t)
%\nonumber\\
%&&
%\times
\int\limits_{\mathbb{R}^{n}} dk\;
\exp\left\{-\frac{1}{4}k^{\mu}M_{\mu\nu}(t)k^{\nu}
+k^{\mu}\mathcal{D}_{\mu}\right\}\;.
\end{eqnarray}
where
\be
J(t)=\det\left(\frac{tiF}{\sinh(tiF)}\right)^{1/2}
\label{25zz}
\ee
and $M(t)$ is a symmetric matrix defined by
\be
M(t)=iF\coth(tiF)\,.
\label{338xzx}
\ee
We would like to stress, at this point, that here and everywhere below all the
functions of the $2$-form $F$ are analytic and should be understood in terms of
a power series in $F$.

Then by using the relation
\begin{equation}
\exp(k^{\mu}\mathcal{D}_{\mu})\delta(u)
=\delta(u+k)\;,
\end{equation}
one obtains
\begin{equation}\label{25}
U_{0}(t|x,x^{\prime})=(4\pi t)^{-n/2}J(t)
\exp\left\{-\frac{1}{4}u^{\mu}M_{\mu\nu}(t)u^{\nu}\right\}\;,
\end{equation}
which is nothing but the Schwinger kernel for an electromagnetic field on
$\RR^n$ \cite{schwinger51}.

To obtain the asymptotic expansion of the heat kernel diagonal we just
need to set $x=x'$ (or $u=0$).
At this point, we notice the following interesting fact.
The operators $t{\cal L}_k$, $tV_k(t\tau)$ and
$T_k(t)$ are differential operators with
homogeneous polynomial coefficients (in $u^\mu$)
of order $\varepsilon^k$.
Recall that $u\sim \varepsilon$, $t\sim \varepsilon^2$ and $F\sim
\varepsilon^{-2}$, so that $tF$ and $Fuu$ are counted as of order zero.
Since the zero order
heat kernel $U_0$ is Gaussian,
then the off-diagonal coefficients $b_k(t|x,x')$ are
polynomials in $u$. The point we want to make now is the following.

\begin{lemma}
The off-diagonal odd-order coefficients $b_{2k+1}$ are odd order polynomials
in $u^\mu$, that is, they satisfy
\be
b_{2k+1}(t|x,x')\Big|_{u\mapsto -u}=-b_{2k+1}(t|x,x')\,,
\ee
and, therefore, vanish on the diagonal,
\be
b^{\rm diag}_{2k+1}(t)=0\,.
\ee
\end{lemma}
\begin{proof}
We discuss the transformation properties of various quantities
under the reflection of the coordinates, $u\mapsto -u$.
First, we note that the operator
${\cal D}$ changes sign, and, therefore, the
operator ${\cal L}_0=-{\cal D}^2$ is invariant.
Next, from the general form of the operator ${\cal L}_k$ discussed above
we see that ${\cal L}_k \mapsto (-1)^k{\cal L}_k$. Therefore,
the same is true for the operator $V_k(t\tau)$, that is,
$V_k \mapsto (-1)^k V_k$.

Now, the operator $T_k(t)$ has the following general form
\be
T_k=t^k\sum_{m=1}^{[k/2]}
\int\limits_0^1 d\tau_1 \cdots \int\limits_0^{\tau_{m-1}} d\tau_m
\sum_{|{\bf j}|=k} C_{m,\, {\bf j}}
V_{j_1}(t\tau_1)\cdots V_{j_m}(t\tau_m)\,,
\ee
where the summation goes over multiindex ${\bf j}=(j_1,\dots,j_m)$
of integers $j_1,\dots,j_m\ge 2$
such that $|{\bf j}|=j_1+\cdots+j_m=k$,
and $C_{m,\, {\bf j}}$ are some numerical
coefficients.
Therefore, the operator $T_k$ transforms as $T_k \mapsto (-1)^k T_k$.

Since the zero-order heat kernel $U_0$ is invariant under the reflection
of coordinates $u\mapsto -u$, we finally find that the coefficients
$b_k$ transform according to $b_k \mapsto (-1)^k b_k$.
Thus, $b_{2k}$ are even polynomials and $b_{2k+1}$ are odd-order
polynomials.

\end{proof}

By using this lemma and by
setting $x=x'$ we obtain the asymptotic expansion of the
heat kernel diagonal
\begin{equation}\label{67}
U^{\rm diag}(t)\sim
(4\pi t)^{-n/2}J(t)\sum_{k=0}^\infty t^k b_{2k}^{\rm diag}(t)
\;,
\end{equation}
where the function $J(t)$ is defined in (\ref{25zz}).
Thus, we obtain
\be
\tilde a_k^{\rm diag}(t)=J(t)b_{2k}^{\rm diag}(t)\,.
\ee

%===========================================================

\subsection{Algebraic Framework}

As we have shown above the evaluation of the heat semigroup is reduced
to the calculation of the operators $V_k(s)$ defined by (\ref{331iga}),
which reduces, in turn, to the computation of general expressions
\begin{equation}
\label{29}
e^{s\mathcal{D}^{2}}u^{\nu_{1}}\cdots u^{\nu_{n}}
\mathcal{D}_{\mu_{1}}\cdots\mathcal{D}_{\mu_{m}}
e^{-s\mathcal{D}^{2}}
=Z^{\nu_{1}}(s)\cdots Z^{\nu_{n}}(s)
A_{\mu_1}(s)\cdots A_{\mu_m}(s)
\;,
\end{equation}
where
\begin{equation}
\label{34}
Z^{\nu}(s)=e^{s\mathcal{D}^{2}}u^{\nu}e^{-s\mathcal{D}^{2}}\;.
\end{equation}
\begin{equation}
\label{31}
A_{\mu}(s)=e^{s\mathcal{D}^{2}}\mathcal{D}_{\mu}e^{-s\mathcal{D}^{2}}\;.
\end{equation}
Obviously, the operators $A_\mu$ and $Z_\nu$ form the algebra
\be
[A_\mu(s),Z^\nu(s)]=\delta^\nu_\mu,\qquad
[A_\mu(s),A_\nu(s)]=iF_{\mu\nu}\,,\qquad
[Z^\mu(s),Z^\nu(s)]=0\,.
\ee

The operators $A_\mu(s)$ and $Z^\nu(s)$ can be computed as follows.
First, we notice that $A_\mu(s)$ satisfies
the differential equation
\begin{equation}
\label{32}
\partial_{s}A_{\mu}(s)=\textrm{Ad}_{\mathcal{D}^{2}}A_{\mu}(s)\;,
\end{equation}
with the initial condition
\begin{displaymath}
A_{\mu}(0)=\mathcal{D}_{\mu}\;.
\end{displaymath}

Hereafter $\textrm{Ad}_{\mathcal{D}^{2}}$ is an operator acting as a commutator,
that is,
\begin{equation}
\textrm{Ad}_{\mathcal{D}^{2}}A_{\mu}(s)
\equiv[\mathcal{D}^{2},A_{\mu}(s)]\;.
\end{equation}

The solution of eq. (\ref{32}) is
\begin{equation}
A_{\mu}(s)
=\exp(s\textrm{Ad}_{\mathcal{D}^{2}})\mathcal{D}_{\mu}\;,
\end{equation}
which can be written in terms of series as
\begin{equation}\label{33}
A_{\mu}(s)=\sum_{k=0}^{\infty}\frac{s^{k}}{k!}
\left(\textrm{Ad}_{\mathcal{D}^{2}}\right)^{k}\mathcal{D}_{\mu}\;.
\end{equation}
Now, by using the algebra (\ref{16a}) we first obtain the commutator
\begin{equation}
[\mathcal{D}^{2},\mathcal{D}_{\mu}]=-2iF_{\mu\alpha}\mathcal{D}^{\alpha}\;,
\end{equation}
and then, by induction,
\begin{equation}
\left(\textrm{Ad}_{\mathcal{D}^{2}}\right)^{k}\mathcal{D}_{\mu}
=(-2i)^{k}F_{\mu\alpha_{1}}F^{\alpha_{1}}{}_{\alpha_{2}}\cdots
F^{\alpha_{k-1}\alpha_{k}}\mathcal{D}_{\alpha_{k}}
=[(-2i F)^{k}]_{\mu\alpha}\mathcal{D}^{\alpha}\;.
\end{equation}
By substituting this result in the series (\ref{33}) we finally find that
\begin{equation}\label{36}
A_{\mu}(s)=\Psi_{\mu}{}^{\alpha}(s)\mathcal{D}_{\alpha}\;,
\end{equation}
where
\begin{equation}\label{35aa}
\Psi(s)=\exp(-2s iF)\;.
\end{equation}

Similarly, for the operators $Z^\nu(s)$ we find
\begin{equation}
Z^{\mu}(s)
=\exp(s\textrm{Ad}_{\mathcal{D}^{2}})u^{\mu}
=\sum_{k=0}^{\infty}\frac{s^{k}}{k!}
\left(\textrm{Ad}_{\mathcal{D}^{2}}\right)^{k}u^{\mu}\;.
\end{equation}
Now, by using the commutators in (\ref{16a}), we find
\begin{equation}
\textrm{Ad}_{\mathcal{D}^{2}}u^{\mu}
=\left[\mathcal{D}^{2},u^{\mu}\right]=2\mathcal{D}^{\mu}\;,
\end{equation}
and then, by induction, we obtain, for $k\geq 2$,
\begin{equation}
\left(\textrm{Ad}_{\mathcal{D}^{2}}\right)^{k}u^{\mu}=2[(-2i F)^{k-1}]^{\mu\alpha}\mathcal{D}_{\alpha}\;.
\end{equation}
Thus the operator $Z^\mu(s)$ in (\ref{31}) takes the form
\begin{equation}
Z^{\mu}(s)
=u^{\mu}-2s\mathcal{D}^{\mu}
+2\sum_{k=2}^{\infty}\frac{s^{k}}{k!}
[(-2i F)^{k-1}]^{\mu\alpha}\mathcal{D}_{\alpha}\;.
\end{equation}
This series can be easily summed up to give
\begin{equation}\label{35a}
Z^{\mu}(s)=u^{\mu}+\Omega^{\mu\alpha}(s)\mathcal{D}_{\alpha}\;,
\end{equation}
where
\begin{equation}\label{35aaa}
\Omega(s)=\frac{1-\exp(-2siF)}{iF}
=2\exp(-siF)\frac{\sinh(siF)}{iF}
\;.
\end{equation}
%The extra factor $t$ is introduced here for convenience.

Now, by using (\ref{35aa}) and (\ref{35aaa}) we obtain
\begin{equation}\label{360}
\Omega^{-1}(s)=\frac{1}{2}iF\left[\coth(siF)+1\right]
=\frac{1}{2}\left[M(s)+iF\right]\;.
\end{equation}
We will need the symmetric and the antisymmetric parts of $\Omega^{-1}(s)$.
By recalling that the matrix $F$ is anti-symmetric it is easy to show
\begin{equation}
\Omega^{-1}_{(\mu\nu)}(s)=\frac{1}{2}M_{\mu\nu}(s)\;.
\end{equation}
\begin{equation}
\Omega^{-1}_{[\mu\nu]}(s)=\frac{1}{2}iF_{\mu\nu}\;,
\label{363iga}
\end{equation}
Here and everywhere below the square brackets denote the complete
antisymmetrization over all indices included.

For the future reference we also notice that
\begin{equation}\label{w}
\Omega^{-1}(s)\Omega^{T}(s)=\Psi^{-1}(s)=\exp(2siF)\;,
\end{equation}
Finally, we define another function
\begin{equation}
\Phi(s)=\Psi(s)\Omega^{-1}(s)
=\left(\Omega^{-1}(s)\right)^T=\frac{1}{2}\left[M(s)-iF\right]\;.
\label{365iga}
\end{equation}
It is useful to remember that the functions $\Psi$,
$F\Omega$ and $\Phi\Omega$ are dimensionless.

%=======================================================

\subsection{Flat Connection}

Next, we transform the operators $Z^\mu$ to define new
(time-dependent) derivative operators by
\bea
D_{\mu}(s)&=&\Omega^{-1}_{\mu\nu}(s)Z^\nu(s)
\;.
\label{36aaz}
\eea
By using the explicit form of the operators $Z^\mu$ and ${\cal D}_\mu$
we have
\bea
D_{\mu}(s)&=&
\mathcal{D}_{\mu}+\Omega^{-1}_{\mu\rho}(s)u^{\rho}
\nonumber\\
&=&\bar{\nabla}_{\mu}+\frac{1}{2}M_{\mu\rho}(s)u^\rho
\;.
\label{36aa}
\eea

Since the operators $Z^\mu$ commute, the operators $D_\mu(s)$ obviously commute
as well. In other words the connection $D_{\mu}$ is flat. Therefore, it can
also be written as
\bea
\label{3600}
D_{\mu}(s)&=&
e^{-\Theta(s)}\bar{\nabla}_{\mu}e^{\Theta(s)}\,,
\eea
where,
\begin{equation}\label{3601}
\Theta(s)=\frac{1}{4}u^{\mu}M_{\mu\nu}(s)u^{\nu}
\end{equation}

Now, we can rewrite the operators
$A_\mu(s)$ and $Z^\mu(s)$
in (\ref{36}) and (\ref{35a}) in terms of the operators
$D_\mu(s)$
\begin{eqnarray}\label{36a}
A_{\mu}(s)&=&\Psi_{\mu}{}^{\alpha}(s)\left(D_{\alpha}(s)
-\Omega^{-1}_{\alpha\rho}(s)u^{\rho}\right)\;,
\nonumber\\
Z^{\mu}(s)&=&\Omega^{\mu\alpha}(s)D_{\alpha}(s)\;.
\end{eqnarray}

%===============================================================
\section{Evaluation of the Operator $T$}
\setcounter{equation}0

The perturbative expansion of the operator $T$ is given by the eq. (\ref{27b}),
with the operators $T_k$ being integrals of the operators $V_k(s)$ and their
product. Thus, according to (\ref{27c})-(\ref{27d}), to compute the operator
$T$ up to the fourth order we need to compute the operators $V_2(s)$, $V_3(s)$,
$V_4(s)$ and $V_2(s_1)V_2(s_2)$.

%========================
\subsection{Second Order}

Now, by using the explicit expression for $\mathscr{L}_{2}$ given by eqs.
(\ref{16}), (\ref{17a}) and (\ref{18b}), utilizing the results of the Section
3, exploiting eqs. (\ref{36a}), (\ref{36b}) and (\ref{36c}), using eqs.
(\ref{35aa}), (\ref{35aaa}), (\ref{w}) and (\ref{365iga}) after some
straightforward but cumbersome calculations we obtain
\begin{eqnarray}
\label{43}
V_{2}(s)
&=&
\frac{1}{6}R
+N_{(2)}^{\sigma}D_{\sigma}
+P_{(2)}^{\gamma\delta}D_{\gamma}D_{\delta}
+W_{(2)}^{\sigma\gamma\delta}D_{\sigma}D_{\gamma}D_{\delta}
\nonumber\\
&+&
Q_{(2)}^{\rho\sigma\gamma\delta}
D_{\rho}D_{\sigma}D_\gamma D_\delta\;,
\end{eqnarray}
where
\begin{eqnarray}
N_{(2)}^{\sigma}
&=&\left(\mathcal{R}^{\mu}{}_{\alpha}
-\frac{1}{3}R^{\mu}{}_{\alpha}\right)
\Omega^{\alpha\sigma}\Phi_{\mu\eta}u^{\eta}\;,\label{44a}
\\
P_{(2)}^{\gamma\delta}&=&
\frac{1}{3}R^{\mu}{}_{\alpha}{}^\nu{}_{\beta}
\Omega^{\alpha(\gamma}\Omega^{|\beta|\delta)}\left[\Phi_{\mu\kappa}
\Phi_{\nu\sigma}u^{\kappa}u^{\sigma}
-\frac{1}{2}M_{\mu\nu}\right]
\nonumber\\
&+&
\frac{1}{24}R^{\nu}{}_{\rho}\Omega^{\rho(\gamma}
\left[\delta_{\nu}{}^{\delta)}
+7\Psi_{\nu}{}^{\delta)}\right]
-\mathcal{R}^{\nu}{}_{\beta}
\Omega^{\beta(\gamma}\Psi_{\nu}{}^{\delta)}\;,
\\
W_{(2)}^{\sigma\delta\gamma}
&=&-\frac{1}{12}
R^{\mu}{}_{\alpha}{}^{\nu}{}_{\beta}\Omega^{\alpha(\sigma}
\Omega^{|\beta|\delta}
\left[\delta_{\nu}{}^{\gamma)}
+7\Psi_{\nu}{}^{\gamma)}\right]\Phi_{\mu\kappa}u^{\kappa}\;,
\\
Q_{(2)}^{\rho\sigma\delta\gamma}&=&
\frac{1}{12}
R^{\mu}{}_{\alpha}{}^{\nu}{}_{\beta}
\Omega^{\alpha(\rho}\Omega^{|\beta|\sigma}
\Psi_{\mu}{}^{\delta}
\left[\delta_{\nu}{}^{\gamma)}
+3\Psi_{\nu}{}^{\gamma)}\right]\;.\label{44b}
\end{eqnarray}
Note that all these coefficients as well as the operators $D_\mu$ depend on the
time variable $s$. We will indicate explicitly the dependence of various
quantities on the time parameter only in the cases when it causes confusion, in
particular, when there are two time parameters.

%=================================================================

\subsection{Third Order}

Similarly, by using the explicit expression for $\mathscr{L}_{3}$ given by
(\ref{16}), (\ref{17aa}) and (\ref{18c}), utilizing the results of the Section
3, exploiting eqs. (\ref{36a}), (\ref{36b}) and (\ref{36c}), using eqs.
(\ref{35aa}), (\ref{35aaa}), (\ref{w}) and (\ref{365iga}) after some
straightforward but cumbersome calculations we obtain
\begin{eqnarray}\label{48a}
V_3(s)
&=&
N_{(3)}^{\sigma}D_{\sigma}+P_{(3)}^{\sigma\rho}D_{\sigma}D_{\rho}
+W_{(3)}^{\sigma\rho\iota}D_{\sigma}D_{\rho}D_{\iota}
\nonumber\\
&&\qquad
+Q_{(3)}^{\sigma\rho\iota\epsilon}D_{\sigma}D_{\rho}D_{\iota}D_{\epsilon}
+Y_{(3)}^{\sigma\rho\iota\epsilon\kappa}D_{\sigma}
D_{\rho}D_{\iota}D_{\epsilon}D_{\kappa}\;,
\end{eqnarray}
where
\bea
\label{48b}
N_{(3)}^{\sigma}
&=&
-\frac{1}{6}\left(\nabla_{\alpha}R
+2\nabla_{\mu}\mathcal{R}^{\mu}{}_{\alpha}\right)\Omega^{\alpha\sigma}\;,
\\[5pt]
P_{(3)}^{\gamma\delta}
&=&
-\frac{1}{6}\left(\nabla^{\mu}R_{\alpha\beta}
-2\nabla_{\alpha}R^{\mu}{}_{\beta}
+4\nabla_{\alpha}\mathcal{R}^{\mu}{}_{\beta}\right)
\Omega^{\alpha(\gamma}\Omega^{|\beta|\delta)}\Phi_{\mu\kappa}u^{\kappa}\;,
\\[5pt]
W_{(3)}^{\sigma\gamma\delta}
&=&
-\frac{1}{6}\nabla_{\alpha}R^{\mu}{}_{\beta}{}^{\nu}{}_{\rho}
\Omega^{\alpha(\sigma}\Omega^{|\beta|\gamma}\Omega^{|\rho|\delta)}
\left[\Phi_{\mu\kappa}\Phi_{\nu\epsilon}u^{\kappa}u^{\epsilon}
-\frac{1}{2}M_{\mu\nu}\right]
\nonumber\\[5pt]
&+&
\frac{1}{6}\left(\nabla^{\mu}R_{\alpha\beta}
-2\nabla_{(\alpha}R^{\mu}{}_{\beta)}
+4\nabla_{(\alpha}\mathcal{R}^{\mu}{}_{\beta)}\right)
\Omega^{\alpha(\sigma}\Omega^{|\beta|\gamma}\Psi_{\mu}{}^{\delta)}\;,
\\[5pt]
Q_{(3)}^{\rho\sigma\gamma\delta}
&=&\frac{1}{3}\nabla_{\alpha}
R^{\mu}{}_{\beta}{}^{\nu}{}_{\epsilon}\Omega^{\alpha(\rho}
\Omega^{|\beta|\sigma}\Omega^{|\epsilon|\gamma}
\Psi_{\nu}^{\delta)}\Phi_{\mu\kappa}u^{\kappa}\;,
\\[5pt]
Y_{(3)}^{\rho\sigma\gamma\delta\epsilon}
&=&
-\frac{1}{6}\nabla_{(\alpha}R^{\mu}{}_{\beta}{}^{\nu}{}_{\eta)}
\Omega^{\alpha(\rho}\Omega^{|\beta|\sigma}\Omega^{|\eta|\gamma}
\Psi_{\mu}^{\delta}\Psi_{\nu}^{\epsilon)}
\;.
\label{48c}
\eea
Here again, for simplicity, we omitted the dependence of the
coefficient functions and the derivatives on the time variable
$s$.

%===============================================================

\subsection{Fourth Order}

\subsubsection{Operator $V_4(s)$}

By taking into account the definition of $\mathscr{L}_{4}$ in (\ref{16}) by
using eqs. (\ref{18d})-(\ref{18a}),  (\ref{36a}), (\ref{36b}) and (\ref{36c}),
and the explicit form of the functions $\Psi$ and $\Omega$, we obtain
\begin{eqnarray}\label{48}
V_4(s)
&=&
P_{(4)}^{\sigma\rho}D_{\sigma}D_{\rho}
+W_{(4)}^{\sigma\rho\iota}D_{\sigma}D_{\rho}D_{\iota}
+Q_{(4)}^{\sigma\rho\iota\epsilon}D_{\sigma}D_{\rho}D_{\iota}D_{\epsilon}
\nonumber\\
&+&Y_{(4)}^{\sigma\rho\iota\epsilon\kappa}
D_{\sigma}D_{\rho}D_{\iota}D_{\epsilon}D_{\kappa}
+S_{(4)}^{\sigma\rho\iota\epsilon\kappa\lambda}
D_{\sigma}D_{\rho}D_{\iota}D_{\epsilon}D_{\kappa}D_{\lambda}
\;,
\end{eqnarray}
where
\begin{eqnarray}
P_{(4)}^{\sigma\rho}
&=&
\frac{1}{60}\left[
R_{\mu\nu}R^{\mu}{}_{\alpha}{}^{\nu}{}_{\beta}
+R_{\mu\nu\lambda\alpha}R^{\mu\nu\lambda}{}_{\beta}
-2R^{\mu}{}_{\alpha}R_{\mu\beta}\right]
\Omega^{\alpha(\rho}\Omega^{|\beta|\sigma)}
\nonumber\\
&+&
\frac{1}{40}\left[\Delta R_{\alpha\beta}
+3\nabla_{\alpha}\nabla_{\beta}R\right]\Omega^{\alpha(\rho}
\Omega^{|\beta|\sigma)}\nonumber\\
&+&
\frac{1}{4}\left[\mathcal{R}_{\mu\alpha}\mathcal{R}^{\mu}{}_{\beta}
+\nabla_{\alpha}\nabla_{\mu}\mathcal{R}^{\mu}{}_{\beta}\right]
\Omega^{\alpha(\rho}\Omega^{|\beta|\sigma)}\;,
\end{eqnarray}
\begin{eqnarray}
W_{(4)}^{\sigma\rho\iota}
&=&\frac{1}{60}
\Big[6\nabla_{\alpha}\nabla^{\mu}R_{\beta\gamma}
+15\nabla_{\alpha}\nabla_{\beta}\mathcal{R}^{\mu}{}_{\gamma}
+15R^{\mu}{}_{\alpha}{}^{\nu}{}_{\beta}\mathcal{R}_{\gamma\nu}
-9\nabla_{\alpha}\nabla_{\beta}R^{\mu}{}_{\gamma}
\nonumber\\
&-&R^{\mu}{}_{\alpha}{}^{\nu}{}_{\beta}R_{\gamma\nu}
-4R^{\mu}{}_{\nu\alpha\lambda}R^{\nu}{}_{\beta}{}^{\lambda}{}_{\gamma}\Big]
\Omega^{\alpha(\sigma}\Omega^{|\beta|\rho}\Omega^{|\gamma|\iota)}
\Phi_{\mu\xi}u^{\xi}\;,
\nonumber\\
\end{eqnarray}
\begin{eqnarray}
Q_{(4)}^{\sigma\rho\iota\epsilon}
&=&\frac{1}{300}
\left[20R^{\mu}{}_{\alpha\lambda\beta}
R^\lambda{}_{\gamma}{}^{\nu}{}_{\delta}
+15\nabla_{\alpha}\nabla_{\beta}
R^{\mu}{}_{\gamma}{}^{\nu}{}_{\delta}\right]
\Omega^{\alpha(\sigma}\Omega^{|\beta|\rho}\Omega^{|\gamma|\iota}
\Omega^{|\delta|\epsilon)}
\nonumber\\
&\times&\left[\Phi_{\mu\xi}\Phi_{\nu\varsigma}u^{\xi}u^{\varsigma}
-\frac{1}{2}M_{\mu\nu}\right]
\nonumber\\
&+&
\frac{1}{240}R^{\alpha}{}_{\nu}R^{\mu\beta\nu\gamma}
\Omega_{\alpha}{}^{(\sigma}\Omega_{\beta}{}^{\rho}\Omega_{\gamma}{}^{\iota}
\left[3\delta_{\mu}{}^{\epsilon)}
+\Psi_{\mu}{}^{\epsilon)}\right]
\nonumber\\
&+&
\frac{1}{240}R_{\lambda}{}^{\mu}{}_{\nu}{}^{\alpha}
R^{\lambda\beta\nu\gamma}
\Omega_{\alpha}{}^{(\sigma}\Omega_{\beta}{}^{\rho}\Omega_{\gamma}{}^{\iota}
\left[3\delta_{\mu}{}^{\epsilon)}
+13\Psi_{\mu}{}^{\epsilon)}\right]
\nonumber\\
&-&
\frac{1}{24}\mathcal{R}_{\nu}{}^{\alpha}
R^{\mu\beta\nu\gamma}
\Omega_{\alpha}{}^{(\sigma}\Omega_{\beta}{}^{\rho}\Omega_{\gamma}{}^{\iota}
\left[\delta_{\mu}{}^{\epsilon)}
+5\Psi_{\mu}{}^{\epsilon)}\right]
\nonumber\\
&+&
\frac{1}{20}\left[3\nabla^{\alpha}\nabla^{\beta}
R^{\mu\gamma}-2\nabla^{\alpha}\nabla^{\mu}R^{\beta\gamma}
-5\nabla^{\alpha}\nabla^{\beta}\mathcal{R}^{\mu\gamma}\right]
\Omega_{\alpha}{}^{(\sigma}\Omega_{\beta}{}^{\rho}\Omega_{\gamma}{}^{\iota}
\Psi_{\mu}{}^{\epsilon)}\;,
\nonumber\\
\end{eqnarray}
%====================
\begin{eqnarray}
Y_{(4)}^{\sigma\rho\iota\epsilon\kappa}
&=&
-\frac{1}{10}\nabla^{\alpha}\nabla^{\beta}
R^{\mu\gamma\nu\delta}
\Omega_{\alpha}{}^{(\sigma}\Omega_{\beta}{}^{\rho}
\Omega_{\gamma}{}^{\iota}
\Omega_{\delta}{}^{\epsilon}
\Psi_{\nu}{}^{\kappa)}\Phi_{\mu\xi}u^{\xi}
\nonumber\\
&-&
\frac{1}{120}R_{\lambda}{}^{\alpha\mu\beta}
R^{\lambda\gamma\nu\delta}
\Omega_{\alpha}{}^{(\sigma}\Omega_{\beta}{}^{\rho}
\Omega_{\gamma}{}^{\iota}
\Omega_{\delta}{}^{\epsilon}
\left[3\delta_{\nu}{}^{\kappa)}
+13\Psi_{\nu}{}^{\kappa)}\right]\Phi_{\mu\xi}u^{\xi}\;,
\nonumber\\
\end{eqnarray}
\begin{eqnarray}
S_{(4)}^{\sigma\rho\iota\epsilon\kappa\lambda}
&=&
\frac{1}{20}\nabla^{\alpha}\nabla^{\beta}
R^{\mu\gamma\nu\delta}\Omega_{\alpha}{}^{(\sigma}
\Omega_{\beta}{}^{\rho}\Omega_{\gamma}{}^{\iota}\Omega_{\delta}{}^{\epsilon}
\Psi_{\mu}{}^{\kappa}\Psi_{\nu}{}^{\lambda)}
\nonumber\\
&+&
\frac{1}{2880}R^{\eta\alpha\mu\beta}
R_{\eta}{}^{\gamma\nu\delta}
\Omega_{\alpha}{}^{(\sigma}\Omega_{\beta}{}^{\rho}
\Omega_{\gamma}{}^{\iota}\Omega_{\delta}{}^{\epsilon}
\Big[62\Psi_{(\mu}{}^{\kappa}\delta_{\nu)}{}^{\lambda)}
\nonumber\\
&+&
125\Psi_{\mu}{}^{\kappa}\Psi_{\nu}{}^{\lambda)}
+5\delta_{\mu}{}^{\kappa}\delta_{\nu}{}^{\lambda)}\Big]\;.
\end{eqnarray}

%=====================================
\subsubsection{Operator $V_2(s_1)V_2(s_2)$}

Next, we need to compute the product of two operators $V_2(s)$ depending on
different times $s_1$ and $s_2$ by using the eq. (\ref{43}). To simplify the
notation we denote the derivatives $D_\mu(s_k)$ depending on different times
$s_k$ simply by $D^{(k)}_\mu$. To present the product $V_2(s_1)V_2(s_2)$ in the
``normal'' form we need to move all derivative operators $D^{(1)}_\mu$ to the
right and all coordinates $u^\nu$ to the left. In order to perform this task we
need the commutator of the derivative operator $D^{(1)}_{\mu}$ with the
coefficients of the operator $V_{2}(s_2)$. First, by using the commutators
listed in Appendix B we obtain the relevant commutators
\begin{eqnarray}
\label{50}
\left[D^{(1)}_{\mu_{1}}\cdots D^{(1)}_{\mu_{n}},
N^{\iota}_{(2)}(s_2)\right]
&=&
n f^{\iota}{}_{(\mu_{1}}(s_2)
D^{(1)}_{\mu_{2}}\cdots D^{(1)}_{\mu_{n})}\;,
\\
\label{51}
\left[D^{(1)}_{\mu_{1}}\cdots D^{(1)}_{\mu_{n}},
P_{(2)}^{\iota\eta}(s_2)\right]
&=&
n(n-1)
g^{\iota\eta}{}_{(\mu_{1}\mu_{2}}(s_2)
D^{(1)}_{\mu_{3}}\cdots D^{(1)}_{\mu_{n})}
\nonumber\\
&&
+nh^{\iota\eta}{}_{(\mu_{1}}(s_2)
D^{(1)}_{\mu_{2}}\cdots D^{(1)}_{\mu_{n})}\;,
%\nonumber\\
%&&
\label{52}
\\
\left[D^{(1)}_{\mu_{1}}\cdots D^{(1)}_{\mu_{n}},
W_{(2)}^{\iota\eta\kappa}(s_2)\right]
&=&
np^{\iota\eta\kappa}{}_{(\mu_{1}}(s_2)
D^{(1)}_{\mu_{2}}\cdots D^{(1)}_{\mu_{n})}\;,
\end{eqnarray}
where
\begin{eqnarray}
f^{\iota}{}_{\lambda}
&=&
\left(\mathcal{R}^{\mu}{}_{\beta}
-\frac{1}{3}R^{\mu}{}_{\beta}\right)\Omega^{\beta\iota}\Phi_{\mu\lambda}\;,
\\
g^{\iota\eta}{}_{\lambda\kappa}
&=&
\frac{1}{3}
R^{\mu}{}_{(\alpha}{}^{\nu}{}_{\beta)}
\Omega^{\alpha\iota}\Omega^{\beta\eta}\Phi_{\mu\lambda}\Phi_{\nu\kappa}\;,
\nonumber\\
h^{\iota\eta}{}_{\lambda}
&=&
\frac{2}{3}R^{\mu}{}_{(\alpha}{}^{\nu}{}_{\beta)}
\Omega^{\alpha\iota}\Omega^{\beta\eta}\Phi_{\mu\kappa}
\Phi_{\nu\lambda}u^{\kappa}\;,
\\
p^{\iota\eta\kappa}{}_{\lambda}
&=&
-\frac{1}{12}R^{\mu}{}_{\alpha}{}^{\nu}{}_{\beta}
\Omega^{\alpha(\iota}\Omega^{|\beta|\eta}
\left[
\delta_{\nu}{}^{\kappa)}
+7\Psi_{\nu}{}^{\kappa)}\right]\Phi_{\mu\lambda}\;.
\end{eqnarray}

Next, by using the expression for the operator $V_2(s)$ in
(\ref{43}) and the non-vanishing
commutators in (\ref{50})-(\ref{52}) we obtain
\begin{eqnarray}\label{53}
V_{2}(s_{1})V_{2}(s_{2})
&=&\frac{1}{36}R^{2}
+\frac{1}{6}R\;\Big[V_{2}(s_{1})
+V_{2}(s_{2})\Big]
+L(s_{1},s_{2})\;,
\end{eqnarray}
where
\begin{equation}\label{53a}
L(s_{1},s_{2})
=\sum_{k=1}^{4}\sum_{n=0}^{4}
C_{(n,k)}^{\mu_{1}\cdots\mu_{n}\nu_{1}\cdots\nu_{k}}(s_{1},s_{2})
D_{\mu_{1}}^{(1)}\cdots D_{\mu_{n}}^{(1)}D_{\nu_{1}}^{(2)}\cdots D_{\nu_{k}}^{(2)}\;,
\end{equation}
and
\begin{eqnarray}
\label{54a}
C_{(0,1)}^{\rho}
&=&
N^{\alpha}_{(2)}(s_{1})f^{\rho}{}_{\alpha}(s_{2})\;,
\nonumber\\
C_{(1,1)}^{\alpha\rho}
&=&
2N^{\alpha}_{(2)}(s_1)N^{\rho}_{(2)}(s_2)
+2P_{(2)}^{\iota\alpha}(s_1)f^{\rho}{}_{\iota}(s_2)\;,
\nonumber\\
C_{(2,1)}^{\alpha\beta\rho}
&=&
2P^{\alpha\beta}_{(2)}(s_1)N^{\rho}_{(2)}(s_2)
+3W_{(2)}^{\kappa\alpha\beta}(s_1)f^{\rho}{}_{\kappa}(s_2)\;,
\nonumber\\
C_{(3,1)}^{\alpha\beta\gamma\rho}
&=&
2W_{(2)}^{\alpha\beta\gamma}(s_1)N^{\rho}_{(2)}(s_2)
+4Q_{(2)}^{\lambda\alpha\beta\gamma}(s_1)f^{\rho}{}_{\lambda}(s_2)\;,
\nonumber\\
C_{(4,1)}^{\alpha\beta\gamma\delta\rho}
&=&
2Q_{(2)}^{\alpha\beta\gamma\delta}(s_1)N^{\rho}_{(2)}(s_2)\;,
\end{eqnarray}
\begin{eqnarray}
C_{(0,2)}^{\rho\sigma}
&=&
N^{\alpha}_{(2)}(s_1)h^{\rho\sigma}{}_{\alpha}(s_2)
+2P_{(2)}^{\alpha\beta}(s_1)g^{\rho\sigma}{}_{\alpha\beta}(s_2)\;,
\nonumber\\
C_{(1,2)}^{\alpha\rho\sigma}
&=&
2N^{\alpha}_{(2)}(s_1)P_{(2)}^{\rho\sigma}(s_2)
+2P_{(2)}^{\alpha\beta}(s_1)h^{\rho\sigma}{}_{\beta}(s_2)
+6W_{(2)}^{\alpha\beta\gamma}(s_1)g^{\rho\sigma}{}_{\beta\gamma}(s_2)\;,
\nonumber\\
C_{(2,2)}^{\alpha\beta\rho\sigma}
&=&
2P^{\alpha\beta}_{(2)}(s_1)P_{(2)}^{\rho\sigma}(s_2)
+3W_{(2)}^{\alpha\beta\gamma}(s_1)h^{\rho\sigma}{}_{\gamma}(s_2)
+12Q_{(2)}^{\alpha\beta\gamma\delta}(s_1)g^{\rho\sigma}{}_{\gamma\delta}(s_2)\;,
\nonumber\\
C_{(3,2)}^{\alpha\beta\gamma\rho\sigma}
&=&
2W^{\alpha\beta\gamma}_{(2)}(s_1)P_{(2)}^{\rho\sigma}(s_2)
+4Q_{(2)}^{\alpha\beta\gamma\delta}(s_1)h^{\rho\sigma}{}_{\delta}(s_2)\;,
\nonumber\\
C_{(4,2)}^{\alpha\beta\gamma\delta\rho\sigma}
&=&
2Q_{(2)}^{\alpha\beta\gamma\delta}(s_1)P_{(2)}^{\rho\sigma}(s_2)\;,
\end{eqnarray}
\begin{eqnarray}
C_{(0,3)}^{\rho\sigma\upsilon}
&=&
N^{\alpha}_{(2)}(s_1)p^{\rho\sigma\upsilon}{}_{\alpha}(s_2)\;,
\nonumber\\
C_{(1,3)}^{\alpha\rho\sigma\upsilon}
&=&
2N^{\alpha}_{(2)}(s_1)W^{\rho\sigma\upsilon}_{(2)}(s_2)
+2P_{(2)}^{\mu\alpha}(s_1)p^{\rho\sigma\upsilon}{}_{\mu}(s_2)\;,
\nonumber\\
C_{(2,3)}^{\alpha\beta\rho\sigma\upsilon}
&=&
2P^{\alpha\beta}_{(2)}(s_1)W^{\rho\sigma\upsilon}_{(2)}(s_2)
+3W_{(2)}^{\mu\alpha\beta}(s_1)p^{\rho\sigma\upsilon}{}_{\mu}(s_2)\;,
\nonumber\\
C_{(3,3)}^{\alpha\beta\gamma\rho\sigma\upsilon}
&=&
2W^{\alpha\beta\gamma}_{(2)}(s_1)W^{\rho\sigma\upsilon}_{(2)}(s_2)
+4Q_{(2)}^{\mu\alpha\beta\gamma}(s_1)p^{\rho\sigma\upsilon}{}_{\mu}(s_2)\;,
\nonumber\\
C_{(4,3)}^{\alpha\beta\gamma\delta\rho\sigma\upsilon}
&=&
2Q^{\alpha\beta\gamma\delta}_{(2)}(s_1)W^{\rho\sigma\upsilon}_{(2)}(s_2)\;,
\end{eqnarray}
\begin{eqnarray}\label{54b}
C_{(1,4)}^{\alpha\rho\sigma\upsilon\chi}
&=&
N^{\alpha}_{(2)}(s_1)Q^{\rho\sigma\upsilon\chi}_{(2)}(s_2)\;,
\nonumber\\
C_{(2,4)}^{\alpha\beta\rho\sigma\upsilon\chi}
&=&
P^{\alpha\beta}_{(2)}(s_1)Q^{\rho\sigma\upsilon\chi}_{(2)}(s_2)\;,
\nonumber\\
C_{(3,4)}^{\alpha\beta\gamma\rho\sigma\upsilon\chi}
&=&
W^{\alpha\beta\gamma}_{(2)}(s_1)Q^{\rho\sigma\upsilon\chi}_{(2)}(s_2)\;,
\nonumber\\
C_{(4,4)}^{\alpha\beta\gamma\delta\rho\sigma\upsilon\chi}
&=&
Q^{\alpha\beta\gamma\delta}_{(2)}(s_1)Q^{\rho\sigma\upsilon\chi}_{(2)}(s_2)\;.
\end{eqnarray}

%====================================================================
\section{Generalized Hermite Polynomials}
\setcounter{equation}0

Thus, we reduced the calculation of the asymptotic expansion of the heat kernel
to the calculation of the derivatives $D_{\mu}(s)$ of the zero order heat
kernel $U_{0}(t|x,x^{\prime})$ given by (\ref{25}). The needed derivatives of
the zero order heat kernel can be expressed in terms of the following symmetric
tensors
\begin{equation}
\label{58}
\mathcal{H}_{\mu_{1}\cdots\mu_{n}}(s)
=U_{0}^{-1}(t|x,x^{\prime})D_{\mu_{1}}(s)\cdots
D_{\mu_{n}}(s)U_{0}(t|x,x^{\prime})\;
\end{equation}
and
\begin{equation}
\label{63b}
\Xi_{\nu_{1}\cdots\nu_{m}\mu_{1}\cdots\mu_{n}}(s_1,s_2)
=U_{0}^{-1}(t|x,x^{\prime})D_{\nu_{1}}^{(1)}
\cdots D_{\nu_{m}}^{(1)}D_{\mu_{1}}^{(2)}\cdots
D_{\mu_{n}}^{(2)}U_{0}(t|x,x^{\prime})\;,
\end{equation}
where we denoted as before $D^{(k)}_\mu=D_\mu(s_k)$.

We recall that the derivatives $D^{(1)}_\mu$ and $D^{(2)}_\nu$ do not commute!
Also, $U_0$ is a scalar function that depends on $x$ and $x'$ only through the
normal coordinates $u^\mu$. The derivative operator $D_\mu(s)$ is defined by
(\ref{36aa}), and, when acting on a scalar function is equal to
\bea
D_\mu(s)&=&\frac{\partial}{\partial u^\mu}
+\frac{1}{2}M_{\mu\nu}(s)u^\nu
\nonumber\\
&=&
e^{-\Theta(s)}\frac{\partial}{\partial u^\mu}e^{\Theta(s)}\,,
\eea
where the tensor $M_{\mu\nu}(s)$ is defined by (\ref{338xzx})
and the function $\Theta(s)$ is a quadratic form defined by (\ref{3601}).

Therefore, by using the explicit form of the zero order heat kernel
(\ref{25}) we see that
the tensors $\mathcal{H}_{\mu_{1}\cdots\mu_{n}}(s)$
can be written in the form
\begin{equation}
\label{58xzx}
\mathcal{H}_{\mu_{1}\cdots\mu_{n}}(s)
=
\exp\{\Theta(t)-\Theta(s)\}
\frac{\partial}{\partial u^{\mu_1}}\cdots
\frac{\partial}{\partial u^{\mu_n}}
\exp\{\Theta(s)-\Theta(t)\}
\;,
\end{equation}

The tensors $\mathcal{H}_{\mu_{1}\cdots\mu_{n}}(s)$ are polynomials in $u^\mu$.
They differ from the usual Hermite polynomials of several variables (see, for
example, \cite{bateman53}) by some normalization. That is why, we call them
just Hermite polynomials. The generating function for Hermite polynomials
\be
{\cal H}(\xi,s)=\sum_{n=0}^\infty \frac{1}{n!}\xi^{\mu_1}
\cdots\xi^{\mu_n}{\cal H}_{\mu_1\dots\mu_n}(s)\,
\ee
can be computed as follows
\bea
{\cal H}(\xi,s)
&=&
\exp{\left\{\Theta(t)-\Theta(s)\right\}}
\exp\left(\xi^\mu\frac{\partial}{\partial u^\mu}\right)
\exp{\left\{\Theta(s)-\Theta(t)\right\}}\;,
\nonumber\\
&=&
\exp\left\{\frac{1}{2}\xi^{\alpha}\Lambda_{\alpha\beta}(s)
\left[\xi^{\beta}+2u^{\sigma}\right]\right\}\;,
\eea
where
\begin{eqnarray}
\label{590}
\Lambda(s)&=&\frac{1}{2}\Big[M(s)-M(t)\Big]
\nonumber\\[5pt]
&=&
\frac{1}{2}\frac{iF}{\sinh(tiF)}\frac{\sinh[(t-s)iF]}{\sinh(siF)}
\;.
\end{eqnarray}

By expanding the exponent in $\xi$ we obtain the Hermite
polynomials explicitly. They can be read off from the expression
\begin{equation}
\label{61z}
\xi^{\mu_{1}}\cdots\xi^{\mu_{n}}\mathcal{H}_{\mu_{1}\cdots\mu_{n}}(s)
=\sum_{k=0}^{\left[\frac{n}{2}\right]}
\frac{(2k)!}{2^{k}k!}{n\choose 2k}
\left(\xi^\alpha\Lambda_{\alpha\beta}(s)\xi^\beta\right)^{k}
\left(\xi^\rho\Lambda_{\rho\sigma}(s)u^\sigma\right)^{n-2k}\;.
\end{equation}
For convenience some low-order Hermite polynomials are given explicitly
in tensorial form in Appendix A.

Similarly, the tensors
$\Xi_{\nu_{1}\cdots\nu_{m}\mu_{1}\cdots\mu_{n}}(s_1,s_2)$ can be written in the
form
\bea
&&\Xi_{\nu_{1}\cdots\nu_{m}\mu_{1}\cdots\mu_{n}}(s_1,s_2)
=
\exp\left[\Theta(t)-\Theta(s_1)\right]
\\[5pt]
&&
\qquad\times
\frac{\partial}{\partial u^{\nu_1}}\cdots
\frac{\partial}{\partial u^{\nu_m}}
\exp\left[\Theta(s_1)-\Theta(s_2)\right]
\frac{\partial}{\partial u^{\mu_1}}\cdots
\frac{\partial}{\partial u^{\mu_n}}
\exp\left[\Theta(s_2)-\Theta(t)\right]
\nonumber
\eea
They are obviously polynomial in $u^\mu$ as well. We call them
Hermite polynomials of second kind.
The generating function for these polynomials is defined by
\begin{equation}\label{63c}
\Xi(\xi,\eta,s_1,s_2)=\sum_{m,n=0}^{\infty}\frac{1}{m!n!}
\xi^{\nu_{1}}\cdots\xi^{\nu_{m}}
\eta^{\mu_{1}}\cdots\eta^{\mu_{n}}\Xi_{\nu_{1}
\cdots\nu_{m}\mu_{1}\cdots\mu_{n}}(s_1,s_2)\;,
\end{equation}
and can be computed as follows
\bea
&&\Xi(\xi,\eta,s_1,s_2)
=\exp\left\{\Theta(t)-\Theta(s_{1})\right\}
\exp\left(\xi^\mu\frac{\partial}{\partial u^\mu}\right)
\exp\left\{\Theta(s_{1})-\Theta(s_{2})\right\}
\nonumber\\
&&\qquad\times
\exp\left(\eta^\nu\frac{\partial}{\partial u^\nu}\right)
\exp\left\{\Theta(s_{2})-\Theta(t)\right\}\;,
\\
&&\qquad
=
\label{63d}
\exp\left\{
\frac{1}{2}\xi^\alpha\Lambda_{\alpha\beta}(s_{1})(\xi^\beta+2u^\beta)
+\frac{1}{2}\eta^\mu\Lambda_{\mu\nu}(s_{2})(\eta^\nu+2u^\nu)
+\xi^\rho\Lambda_{\rho\sigma}(s_{2})\eta^\sigma\right\}\;,
\nonumber
\eea

Notice that
\be
\Xi(\xi,\eta,s_1,s_2)={\cal H}(\xi,s_1){\cal H}(\eta,s_2)
\exp\left\{
\xi^\rho\Lambda_{\rho\sigma}(s_{2})\eta^\sigma\right\}
\ee
This enables one to express all Hermite polynomials of second kind
$\Xi_{(n)}(s_1,s_2)$
in terms of the Hermite polynomials
${\cal H}_{(m)}(s_1)$, ${\cal H}_{(l)}(s_2)$, and the matrix
$\Lambda(s_2)$.
Namely, they can be read off from the expression
\bea
&&
\xi^{\nu_{1}}\cdots\xi^{\nu_{m}}
\eta^{\mu_{1}}\cdots\eta^{\mu_{n}}\Xi_{\nu_{1}
\cdots\nu_{m}\mu_{1}\cdots\mu_{n}}(s_1,s_2)
=
\sum_{k=0}^{\min(m,n)}
k!{m\choose k}{n\choose k}
\label{613iga}
\\
&&
\qquad
\times
\xi^{\nu_{1}}\cdots\xi^{\nu_{m-k}}\mathcal{H}_{\nu_{1}\cdots\nu_{m-k}}(s_1)
\eta^{\mu_{1}}\cdots\eta^{\mu_{n-k}}\mathcal{H}_{\mu_{1}\cdots\mu_{n-k}}(s_2)
\left(\xi^\rho\Lambda_{\rho\sigma}(s_2)\eta^\sigma\right)^{k}
\nonumber
\eea

%=========================================
\section{Off-diagonal Coefficients $b_k$}
\setcounter{equation}0

By using the machinery developed above, we can now write the coefficients of
the asymptotic expansion of the heat kernel in terms of generalized Hermite
polynomials. We define the following quantity
\bea
b_{2, (1)}(t|x,x')
&=&
\int\limits_{0}^{1}d\tau\Big[N_{(2)}^{\sigma}(t\tau)\mathcal{H}_{\sigma}(t\tau)
+P_{(2)}^{\gamma\delta}(t\tau)\mathcal{H}_{\gamma\delta}(t\tau)
+W_{(2)}^{\sigma\gamma\delta}(t\tau)\mathcal{H}_{\sigma\gamma\delta}(t\tau)
\nonumber\\
&&
+Q_{(2)}^{\rho\sigma\gamma\delta}(t\tau)
\mathcal{H}_{\rho\sigma\gamma\delta}(t\tau)\Big]\;.
\eea
Then, by referring to the formulas (\ref{43}), (\ref{48a}), (\ref{48}) and
(\ref{53}) and by using the following formula for multiple integrals
\begin{equation}
\int\limits_{a}^{b}d\tau_{n}
\int\limits_{a}^{\tau_{n}}d\tau_{n-1}
\cdots\int\limits_{a}^{\tau_{2}}d\tau_1 f(\tau_1)
=\frac{1}{(n-1)!}\int\limits_{a}^{b}d\tau\;(b-\tau)^{n-1}f(\tau)\,,
\end{equation}
we obtain
\bea
\label{63}
b_{2}(t|x,x^{\prime})
&=&\frac{1}{6}R+b_{2, (1)}(t|x,x')\,,
\eea
\begin{eqnarray}\label{63a}
b_{3}(t|x,x^{\prime})
&=&t^{-1/2}\int\limits_{0}^{1}d\tau\Big[
N_{(3)}^{\sigma}(t\tau)\mathcal{H}_{\sigma}(t\tau)
+P_{(3)}^{\gamma\delta}(t\tau)\mathcal{H}_{\gamma\delta}(t\tau)
+W_{(3)}^{\sigma\gamma\delta}(t\tau)\mathcal{H}_{\sigma\gamma\delta}(t\tau)
\nonumber\\
&&+
Q_{(3)}^{\rho\sigma\gamma\delta}(t\tau)
\mathcal{H}_{\rho\sigma\gamma\delta}(t\tau)
+Y_{(3)}^{\iota\rho\sigma\gamma\delta}(t\tau)
\mathcal{H}_{\iota\rho\sigma\gamma\delta}(t\tau)\Big]\;.
\end{eqnarray}
\begin{eqnarray}
\label{64}
b_{4}(t|x,x^{\prime})
&=&
\frac{1}{72}R^{2}
+\frac{1}{6}R b_{2, (1)}(t|x,x')
\nonumber\\
&&
+t^{-1}\int\limits_{0}^{1}d\tau\Big[
P^{\iota\epsilon}_{(4)}(t\tau)\mathcal{H}_{\iota\epsilon}(t\tau)
+W^{\iota\epsilon\kappa}_{(4)}(t\tau)
\mathcal{H}_{\iota\epsilon\kappa}(t\tau)
\nonumber\\
&&
+Q^{\iota\epsilon\kappa\lambda}_{(4)}(t\tau)
\mathcal{H}_{\iota\epsilon\kappa\lambda}(t\tau)
+Y^{\iota\epsilon\kappa\lambda\eta}_{(4)}(t\tau)
\mathcal{H}_{\iota\epsilon\kappa\lambda\eta}(t\tau)
+S^{\iota\epsilon\kappa\lambda\eta\gamma}_{(4)}(t\tau)
\mathcal{H}_{\iota\epsilon\kappa\lambda\eta\gamma}(t\tau)\Big]
\nonumber\\
&&+
\sum_{k=1}^{4}\sum_{n=0}^{4}\int\limits_{0}^{1}d\tau_{2}
\int\limits_{0}^{\tau_{2}}d\tau_{1}\;
C_{(n,k)}^{\mu_{1}\cdots\mu_{n}\nu_{1}\cdots\nu_{k}}(t\tau_{1},t\tau_{2})
\Xi_{\mu_{1}\cdots\mu_{n}\nu_{1}\cdots\nu_{k}}(t\tau_{1},t\tau_{2})\;.
\nonumber\\
&&
\end{eqnarray}

%=====================================================
\section{Diagonal Coefficients $b_k$}
\setcounter{equation}0

In order to obtain the diagonal values $b_k^{\rm diag}(t)$ of the coefficients
$b_k(t|x,x')$ we just need to set $u=0$ in eqs. (\ref{63}), (\ref{63a}) and
(\ref{64}). For the rest of this section we will employ the usual convention of
denoting the coincidence limit by square brackets, that is,
\begin{equation}
\left[f(u)\right]^{\rm diag}=f(0).
\end{equation}

By inspection of the equation defining the generalized Hermite polynomials in
Appendix A one can easily notice that, in the coincidence limit, all the ones
with an odd number of indices vanish identically, namely
\begin{equation}
\label{69}
\left[\mathcal{H}_{\mu_{1}\cdots\mu_{2n+1}}\right]^{\rm diag}=0\;.
\end{equation}
By using the last remark we have the following expression for the
coincidence limit of (\ref{63}), i.e.
\begin{equation}
\label{70}
b^{\rm diag}_{2}(t)
=\frac{1}{6}R
+b^{\rm diag}_{2, (1)}(t)\,,
\ee
where
\be
b^{\rm diag}_{2, (1)}(t)
=\int\limits_{0}^{1}d\tau\left[P_{(2)}^{\gamma\delta}(t\tau)
\mathcal{H}_{\gamma\delta}(t\tau)
+Q_{(2)}^{\rho\sigma\gamma\delta}(t\tau)
\mathcal{H}_{\rho\sigma\gamma\delta}(t\tau)\right]^{\rm diag}\;.
\end{equation}
By using the explicit form of the coefficients $P_{(2)}$,
$Q_{(2)}$ and the generalized Hermite polynomials in
Appendix A,
we obtain
\begin{equation}
b^{\rm diag}_{2, (1)}(t)
=J_{(1)}{}^{\alpha\beta}{}_{\mu\nu}(t)
R^{\mu}{}_{\alpha}{}^{\nu}{}_{\beta}
+J_{(2)}{}^{\mu\nu}(t)R_{\mu\nu}
+J_{(3)}{}^{\mu\nu}(t)\mathcal{R}_{\mu\nu}\;,
\end{equation}
where
\begin{eqnarray}
J_{(1)}{}^{\alpha\beta}{}_{\mu\nu}(t)
&=&
\int\limits_{0}^{1}d\tau\;
\Bigg\{-\frac{1}{6}
\Omega^{\alpha\gamma}\Omega^{\beta\delta}
M_{\mu\nu}\Lambda_{\gamma\delta}
\nonumber\\
&&
+\frac{1}{4}\left(
\delta_{\nu}{}^{\gamma}
+3\Psi_{\nu}{}^{\gamma}\right)
\Omega^{\alpha\rho}\Omega^{\beta\sigma}
\Psi_{\mu}{}^{\delta}
\Lambda_{(\rho\sigma}\Lambda_{\delta\gamma)}\Bigg\}\;,
\\
J_{(2)}{}^{\mu\nu}(t)
&=&
\frac{1}{24}\int\limits_{0}^{1}d\tau
\left(\delta^{(\nu}{}_{\delta}
+7\Psi^{(\nu}{}_{\delta}\right)
\Omega^{\mu)\gamma}
\Lambda_{\gamma}{}^{\delta}\;,
\\
J_{(3)}{}^{\mu\nu}(t)
&=&
\int\limits_{0}^{1}d\tau\;
\Omega^{[\mu}{}_{\gamma}\Psi^{\nu]\delta}\Lambda^{\gamma}{}_{\delta}\;.
\label{70b}
\end{eqnarray}
Here all functions in the integrals depend on $t\tau$.

Next, we introduce the following matrices
\begin{equation}
\label{71}
\mathcal{A}(s)
=\Omega(s)\Lambda(s)
=\frac{1}{2}\frac{\exp[(t-2s)iF]-\exp(-tiF)}{\sinh(tiF)}\;.
\end{equation}
\begin{equation}\label{71b}
\mathcal{B}(s)
=\Omega(s)\Lambda(s)\Omega(s)^{T}
=\frac{\coth(tiF)}{iF}
-\frac{\cosh[(t-2s)iF]}{iF\sinh(tiF)}\;,
\end{equation}
\bea
\Gamma(s)
&=&\Omega^{-1}(s)
-\frac{1}{4}\Psi(s)\Lambda(s)
-\frac{3}{4}\Lambda(s)
\nonumber\\
&&
=\frac{1}{8}\left(3iF\coth(tiF)
+\frac{iF}{\sinh(tiF)}\cosh[(t-2s)iF]\right)\;.
\eea
Then, by using the relation
\begin{equation}\label{71a}
\Omega(s)\Lambda(s)\Psi(s)^{T}
=\Omega^{T}(s)\Lambda(s)=\mathcal{A}^{T}(s)\;
\end{equation}
we obtain
\bea
\label{72c}
J_{(1)}{}^{\alpha\beta}{}_{\mu\nu}(t)
&=&\int\limits_{0}^{1}d\tau
\Bigg\{
-\frac{1}{3}\mathcal{B}^{\alpha\beta}(t\tau)\Gamma_{(\mu\nu)}(t\tau)
\nonumber\\
&&
+\frac{1}{6}\left(\mathcal{A}_{\mu}{}^{(\alpha}(t\tau)
\mathcal{A}^{\beta)}{}_{\nu}(t\tau)
+3\mathcal{A}_{(\mu}{}^{\alpha}(t\tau)
\mathcal{A}_{\nu)}{}^{\beta}(t\tau)\right)\Bigg\}\;,
\\
\label{72a}
J_{(2)}{}^{\mu\nu}(t)
&=&\frac{1}{3}\int\limits_{0}^{1}
d\tau\mathcal{A}^{(\mu\nu)}(t\tau)
=\frac{1}{6}\delta^{\mu\nu}\;,
%\end{equation}
%\begin{equation}
\\
\label{72b}
J_{(3)}{}^{\mu\nu}(t)
&=&-\int\limits_{0}^{1}d\tau\mathcal{A}^{[\mu\nu]}(t\tau)
=
-\frac{1}{2}\left(\frac{1}{tiF}-\coth(tiF)\right)^{[\mu\nu]}\;.
\eea
Unfortunately the integral $J_{(1)}{}^{\alpha\beta}{}_{\mu\nu}$
can not be computed explicitly, in general.

As we already mentioned above all odd order coefficients $b_{2k+1}$ have zero
diagonal values. We see this directly for the coefficient $b_3$, which is given
by (\ref{63a}). That is, by recalling the formulas in (\ref{48b}) through
(\ref{48c}) and the remark (\ref{69}) we have
\begin{equation}
b^{\rm diag}_{3}(t)=0\;.
\end{equation}

Finally, we evaluate the diagonal values of fourth order coefficient $b_4$
given by (\ref{64}). It can be written as follows
\begin{equation}\label{72}
b^{\rm diag}_{4}(t)
=\frac{1}{72}R^{2}
+\frac{1}{6}Rb^{\rm diag}_{2, (1)}(t)
+b^{\rm diag}_{4, (2)}(t)
+b^{\rm diag}_{4, (3)}(t)\;.
\end{equation}
By noticing that for odd $n+k$, the diagonal values of the coefficients
$C_{(n,k)}$ vanish,
\begin{equation}
\left[C_{(n,k)}^{\mu_{1}\cdots\mu_{n}\nu_{1}\cdots\nu_{k}}
\right]^{\rm diag}=0\;,
\end{equation}
and by using the explicit form of Hermite polynomials and the generating
function (\ref{63d}) we obtain
\begin{eqnarray}\label{74}
b^{\rm diag}_{4, (2)}(t)
&=&t^{-1}\int\limits_{0}^{1}d\tau
\bigg\{P^{\iota\epsilon}_{(4)}(t\tau)\Lambda_{\iota\epsilon}(t\tau)
+3\left[Q^{\iota\epsilon\kappa\lambda}_{(4)}(t\tau)\right]^{\rm diag}
\Lambda_{(\iota\epsilon}(t\tau)\Lambda_{\kappa\lambda)}(t\tau)
\nonumber\\
&&
+15S^{\iota\epsilon\kappa\lambda\eta\gamma}_{(4)}(t\tau)
\Lambda_{(\iota\epsilon}\Lambda_{\kappa\lambda}(t\tau)
\Lambda_{\eta\gamma)}(t\tau)\bigg\}\;,
\end{eqnarray}
\begin{eqnarray}\label{75}
b^{\rm diag}_{4, (3)}(t)
&=&
\int\limits_{0}^{1}d\tau_{2}\int\limits_{0}^{\tau_{2}}d\tau_{1}
\bigg\{2\Big[P_{(2)}^{\iota\alpha}(\tau_{1})\Big]^{\rm diag}
f^{\rho}{}_{\iota}(\tau_{2})\Lambda^{(2)}_{\alpha\rho}
+2\Big[P_{(2)}^{\alpha\beta}(\tau_{1})\Big]^{\rm diag}
g^{(\rho\sigma)}{}_{\alpha\beta}(\tau_{2})\Lambda^{(2)}_{\rho\sigma}
\nonumber\\
&+&
12Q_{(2)}^{\lambda\alpha\beta\gamma}(\tau_{1})
f^{\rho}{}_{\lambda}(\tau_{2})
\Lambda^{(1)}_{\alpha\beta}\Lambda^{(2)}_{\gamma\rho}
+\Big(2\Big[P^{\alpha\beta}_{(2)}(\tau_{1})\Big]^{\rm diag}
\Big[P_{(2)}^{\rho\sigma}(\tau_{2})\Big]^{\rm diag}
\nonumber\\[4pt]
&+&
12Q_{(2)}^{\alpha\beta\gamma\delta}(\tau_{1})
g^{(\rho\sigma)}{}_{\gamma\delta}(\tau_{2})\Big)
\left(\Lambda^{(1)}_{\alpha\beta}\Lambda^{(2)}_{\rho\sigma}
+2\Lambda^{(2)}_{\alpha\rho}\Lambda^{(2)}_{\beta\sigma}\right)
\nonumber\\[4pt]
&+&
6\Big[P_{(2)}^{\mu\alpha}(\tau_{1})\Big]^{\rm diag}
p^{(\rho\sigma\nu)}{}_{\mu}(\tau_{2})
\Lambda^{(2)}_{\alpha\rho}\Lambda^{(2)}_{\sigma\nu}
\nonumber\\[4pt]
&+&
2Q_{(2)}^{\alpha\beta\gamma\delta}(\tau_{1})
\Big[P_{(2)}^{\rho\sigma}(\tau_{2})\Big]^{\rm diag}
\left(3\Lambda^{(1)}_{\alpha\beta}
\Lambda^{(1)}_{\gamma\delta}\Lambda^{(2)}_{\rho\sigma}
+12\Lambda^{(1)}_{\alpha\beta}
\Lambda^{(2)}_{\gamma\rho}\Lambda^{(2)}_{\delta\sigma}\right)
\nonumber\\[4pt]
&+&
4Q_{(2)}^{\mu\alpha\beta\gamma}(\tau_{1})
p^{(\rho\sigma\nu)}{}_{\mu}(\tau_{2})
\left(9\Lambda^{(1)}_{\alpha\beta}
\Lambda^{(2)}_{\gamma\rho}
\Lambda^{(2)}_{\sigma\nu}
+6\Lambda^{(2)}_{\alpha\rho}
\Lambda^{(2)}_{\beta\sigma}\Lambda^{(2)}_{\gamma\nu}\right)
\nonumber\\[4pt]
&+&
\Big[P^{\alpha\beta}_{(2)}(\tau_{1})\Big]^{\rm diag}
Q^{\rho\sigma\nu\chi}_{(2)}(\tau_{2})
\left(3\Lambda^{(1)}_{\alpha\beta}
\Lambda^{(2)}_{\rho\sigma}\Lambda^{(2)}_{\nu\chi}
+12\Lambda^{(2)}_{\alpha\rho}
\Lambda^{(2)}_{\beta\sigma}\Lambda^{(2)}_{\nu\chi}\right)
\nonumber\\[4pt]
&+&
Q^{\alpha\beta\gamma\delta}_{(2)}(\tau_{1})
Q^{\rho\sigma\nu\chi}_{(2)}(\tau_{2})
\Big(9\Lambda^{(1)}_{\alpha\beta}
\Lambda^{(1)}_{\gamma\delta}\Lambda^{(2)}_{\rho\sigma}
\Lambda^{(2)}_{\nu\chi}
+72\Lambda^{(1)}_{\alpha\beta}\Lambda^{(2)}_{\rho\gamma}
\Lambda^{(2)}_{\sigma\delta}
\Lambda^{(2)}_{\nu\chi}
\nonumber\\[4pt]
&+&
24\Lambda^{(2)}_{\alpha\rho}
\Lambda^{(2)}_{\beta\sigma}
\Lambda^{(2)}_{\gamma\nu}\Lambda^{(2)}_{\delta\chi}\Big)\bigg\}\;,
\end{eqnarray}
where the superscript on the matrix $\Lambda$ denotes its dependence on
either $t\tau_{1}$ or $t\tau_{2}$.

We see that the scalar curvature appears only in the term $b^{\rm diag}_{2,
(1)}(t)$. Now, the term $b^{\rm diag}_{4, (2)}(t)$ only contains derivatives of
the curvature and quantities which are quadratic in the curvature with some of
their indices contracted. It has the following form
\begin{eqnarray}
b^{\rm diag}_{4, (2)}(t)
&=&\frac{1}{60}
B_{\alpha\beta}(t)R_{\mu\nu\lambda}{}^{\alpha}
R^{\mu\nu\lambda\beta}
+\mathrm{A}^{(1)}_{\lambda\alpha\gamma\beta}(t)
R_{\mu}{}^{\lambda}{}_{\nu}{}^{\alpha}
R^{\mu\gamma\nu\beta}
+\mathrm{A}^{(2)}_{\alpha\mu\beta\gamma\nu\delta}(t)
R_{\eta}{}^{\alpha\mu\beta}
R^{\eta\gamma\nu\delta}
\nonumber\\
&+&\frac{1}{60}
B_{\alpha\beta}(t)R_{\mu\nu}
R^{\mu\alpha\nu\beta}
+\mathrm{A}^{(3)}_{\alpha\mu\beta\gamma}(t)R^{\alpha}{}_{\nu}
R^{\mu\beta\nu\gamma}
-\frac{1}{30}B_{\alpha\beta}(t)R_{\mu}{}^{\alpha}R^{\mu\beta}
\nonumber\\
&+&
\mathrm{A}^{(4)}_{\alpha\mu\beta\gamma}(t)
\mathcal{R}_{\nu}{}^{\alpha}R^{\mu\beta\nu\gamma}
+\frac{1}{4}B_{\alpha\beta}(t)\mathcal{R}_{\nu}{}^{\alpha}
\mathcal{R}^{\nu\beta}
\nonumber\\
&+&
\mathrm{A}^{(5)}_{\alpha\beta\mu\gamma\nu\delta}(t)
\nabla^{\alpha}\nabla^{\beta}R^{\mu\gamma\nu\delta}
+
\mathrm{A}^{(6)}_{\alpha\beta\mu\nu}(t)
\nabla^{\alpha}\nabla^{\beta}R^{\mu\nu}
+\frac{1}{40}B_{\alpha\beta}(t)\Delta R^{\alpha\beta}
\nonumber\\
&+&
\frac{3}{40}B_{\alpha\beta}(t)\nabla^{\alpha}\nabla^{\beta}R\;
+\mathrm{A}^{(7)}_{\alpha\beta\mu\nu}(t)
\nabla^{\alpha}\nabla^{\beta}\mathcal{R}^{\mu\nu}
+\frac{1}{4}B_{\alpha\beta}(t)\nabla^{\alpha}\nabla_{\mu}
\mathcal{R}^{\mu\beta}\;,
\end{eqnarray}
Here the tensors $a^{(i)}(s)$ are functions that only depend
on $F$ (but not on the Riemann curvature) defined by
\begin{eqnarray}
a^{(1)}{}_{\lambda\alpha\gamma\beta}(s)
&=&
\frac{3}{80}\mathcal{B}_{(\alpha\gamma}\mathcal{A}_{\beta)\lambda}
+\frac{13}{80}\mathcal{B}_{(\alpha\gamma}\mathcal{A}_{|\lambda|\beta)}\;,
\\
a^{(2)}{}_{\alpha}{}^{\mu}{}_{\beta\gamma}{}^{\nu}{}_{\delta}(s)
&=&
\frac{1}{480}\mathcal{B}_{\alpha(\beta}\mathcal{B}_{\gamma\delta)}
\left(31(\Psi\Lambda)^{(\mu\nu)}
+65\Lambda^{\mu\nu}\right)
-\frac{1}{10}M^{\mu\nu}\mathcal{B}_{\alpha(\beta}
\mathcal{B}_{\gamma\delta)}
\nonumber\\
&+&\frac{187}{480}\mathcal{B}_{\alpha(\beta}
\mathcal{A}_{\gamma}{}^{(\mu}\mathcal{A}^{\nu)}{}_{\delta)}
+\frac{31}{240}\mathcal{B}_{(\beta\gamma}
\mathcal{A}^{(\nu}{}_{\delta)}\mathcal{A}_{\alpha}{}^{\mu)}
\nonumber\\
&+&\frac{25}{96}
\mathcal{B}_{(\beta\gamma}
\mathcal{A}^{(\nu}{}_{\delta)}\mathcal{A}^{\mu)}{}_{\alpha}
+\frac{1}{96}\left(\mathcal{B}_{\alpha(\beta}
\mathcal{A}_{\gamma}{}^{(\mu}\mathcal{A}_{\delta)}{}^{\nu)}
+\mathcal{B}_{(\beta\gamma}
\mathcal{A}_{\delta)}{}^{(\nu}\mathcal{A}_{\alpha}{}^{\mu)}\right)\;,\;\;\;\;\;\;\;\;\;\;
\\
a^{(3)}{}_{\alpha\mu\beta\gamma}(s)
&=&\frac{3}{80}\mathcal{B}_{(\alpha\beta}\mathcal{A}_{\gamma)\mu}
+\frac{1}{80}\mathcal{B}_{(\alpha\beta}\mathcal{A}_{|\mu|\gamma)}\;,
\\
a^{(4)}{}_{\alpha\mu\beta\gamma}(s)
&=&
-\frac{1}{8}\mathcal{B}_{(\alpha\beta}
\mathcal{A}_{\gamma)\mu}
-\frac{5}{8}\mathcal{B}_{(\alpha\beta}\mathcal{A}_{|\mu|\gamma)}\;,
\\
a^{(5)}{}_{\alpha\beta}{}^{\mu}{}_{\gamma}{}^{\nu}{}_{\delta}(s)
&=&
-\frac{3}{40}\mathcal{B}_{\alpha(\beta}
\mathcal{B}_{\gamma\delta)}M^{\mu\nu}(t)
+\frac{3}{10}\mathcal{B}_{\alpha(\beta}
\mathcal{A}^{(\mu}{}_{\gamma}\mathcal{A}^{\nu)}{}_{\delta)}
\nonumber\\
&+&\frac{3}{10}\mathcal{B}_{(\beta\gamma}
\mathcal{A}^{(\mu}{}_{\delta)}\mathcal{A}^{\nu)}{}_{\alpha}\;,
\\
a^{(6)}{}_{\alpha\beta\mu\nu}(s)
&=&\frac{9}{20}\mathcal{B}_{(\alpha\beta}
\mathcal{A}_{|\mu|\nu)}
-\frac{3}{10}\mathcal{B}_{\mu(\alpha}
\mathcal{A}_{\beta\nu)}\;,\\
a^{(7)}{}_{\alpha\beta\mu\nu}(s)
&=&
-\frac{3}{4}\mathcal{B}_{(\alpha\beta}\mathcal{A}_{|\mu|\nu)}\;.
\end{eqnarray}
All functions here are evaluated at the time $s$
(unless  specified otherwise).

The term $b^{\rm diag}_{4, (3)}(t)$ only contains quantities which are
quadratic in the curvature with none of their indices contracted.
It has the form
\begin{eqnarray}
b^{\rm diag}_{4, (3)}(t)&=&
\mathrm{D}^{(1)}_{\alpha\beta\mu\nu\gamma\delta\rho\sigma}(t)
R^{\alpha\beta\mu\nu}R^{\gamma\delta\rho\sigma}
+
\mathrm{D}^{(2)}_{\mu\nu\alpha\beta\rho\sigma}(t)
R^{\mu\nu}R^{\alpha\beta\rho\sigma}
+\mathrm{D}^{(3)}_{\mu\nu\alpha\beta}(t)
R^{\mu\nu}R^{\alpha\beta}
\nonumber\\
&+&
\mathrm{D}^{(4)}_{\mu\nu\alpha\beta\rho\sigma}(t)
\mathcal{R}^{\mu\nu}R^{\alpha\beta\rho\sigma}
+
\mathrm{D}^{(5)}_{\mu\nu\alpha\beta}(t)
\mathcal{R}^{\mu\nu}R^{\alpha\beta}
+
\mathrm{D}^{(6)}_{\mu\nu\alpha\beta}(t)
\mathcal{R}^{\mu\nu}\mathcal{R}^{\alpha\beta}
\;,
\end{eqnarray}
where $\mathrm{D}^{(i)}_{\mu_{1}\cdots\mu_{n}}(t)$ are some tensor-valued
functions that depend on $tF$.
They have the form
\begin{equation}
\mathrm{D}^{(i)}_{\mu_{1}\cdots\mu_{n}}(t)
=\int\limits_{0}^{1}d\tau_{2}\int\limits_{0}^{\tau_{2}}
d\tau_{1}\;d^{(i)}_{\mu_{1}\cdots\mu_{n}}(t\tau_{1},t\tau_{2})\;.
\end{equation}

%========================================================

To describe our results for the tensors $d^{(k)}$
we define new tensors
\begin{eqnarray}
\mathcal{E}_{(p)\mu}{}^{\nu}
&=&
\delta_{\mu}{}^{\nu}+p\;\Psi_{\mu}{}^{\nu}\;,
\\
\label{76}
\mathcal{S}_{\alpha\beta\rho\sigma\iota\kappa}
&=&
\mathcal{B}_{\beta\sigma}\Phi_{\alpha\iota}
\Phi_{\rho\kappa}
-\mathcal{A}_{\beta(\iota}\mathcal{A}_{|\sigma|\kappa)}
M_{\alpha\rho}
\nonumber\\
&-&\frac{3}{4}\Omega_{\beta}{}^{(\eta}\Omega
_{\sigma}{}^{\chi}{}\mathcal{E}_{(1)\rho}{}^{\epsilon)}
\Phi_{\alpha\iota}\Lambda_{\kappa\eta}\Lambda_{\chi\epsilon}
%\nonumber\\
%&+&
+\frac{3}{2}\Omega_{\beta}{}^{(\epsilon}
\Omega_{\sigma}{}^{\lambda}\Psi_{\alpha}{}^{\eta}\mathcal{E}_{(3)\rho}{}^{\chi)}
\Lambda_{\iota\epsilon}\Lambda_{\kappa\lambda}
\Lambda_{\eta\chi}\;,\;\;\;\;
\end{eqnarray}
\begin{eqnarray}
\mathcal{V}_{\gamma\delta\rho\sigma\iota\kappa\eta\chi}
(t\tau_{1},t\tau_{2})
&=&
\Lambda_{\eta\chi}(t\tau_{1})
\Big(\mathcal{B}_{\delta\sigma}\Phi_{\gamma\iota}
\Phi_{\rho\kappa}\Big)(t\tau_{2})
+2\Big(\mathcal{A}_{\delta\iota}\mathcal{A}_{\sigma\kappa}
\Phi_{\gamma\eta}\Phi_{\rho\chi}\Big)(t\tau_{2})
\nonumber\\
&-&\frac{1}{4}\Big(\Lambda_{\iota\kappa}\Lambda_{\eta\chi}\Big)
(t\tau_{1})\Big(\mathcal{B}_{\delta\sigma}M_{\gamma\rho}\Big)
(t\tau_{2})
\nonumber\\
&-&\Lambda_{\iota\kappa}(t\tau_{1})
\Big(\mathcal{A}_{\delta(\chi}\mathcal{A}_{|\sigma|\eta)}
M_{\gamma\rho}\Big)(t\tau_{2})
\nonumber\\
&-&\frac{3}{4}\left\{\Lambda_{\kappa\eta}(t\tau_{1})
\Big(\Lambda_{\chi\epsilon}\Lambda_{\omega\tau}\big)
(t\tau_{2})+\frac{2}{3}\Lambda_{\kappa\epsilon}(t\tau_{1})
\Big(\Lambda_{\eta\omega}
\Lambda_{\chi\tau}\Big)(t\tau_{2})\right\}
\nonumber\\
&\times&\Big(\Omega_{\delta}{}^{(\epsilon}\Omega_{\sigma}
{}^{\omega}{}\mathcal{E}_{(1)\rho}{}^{\tau)}\Phi_{\gamma\iota}\Big)(t\tau_{2})
\nonumber\\
&+&\frac{3}{16}\Bigg\{\Big(\Lambda_{\iota\kappa}
\Lambda_{\eta\chi}\Big)(t\tau_{1})
\Big(\Lambda_{\epsilon\tau}\Lambda_{\omega\lambda}\Big)(t\tau_{2})
+8\Lambda_{\iota\kappa}(t\tau_{1})\Big(\Lambda_{\epsilon\eta}
\Lambda_{\tau\chi}\Lambda_{\omega\lambda}\Big)(t\tau_{2})
\nonumber\\
&+&\frac{8}{3}\Big(\Lambda_{\iota\epsilon}
\Lambda_{\kappa\tau}\Lambda_{\eta\omega}
\Lambda_{\chi\lambda}\Big)(t\tau_{2})\Bigg\}
\left(\Omega_{\delta}{}^{(\epsilon}\Omega_{\sigma}{}^{\tau}
\Psi_{\gamma}{}^{\omega}\mathcal{E}_{(3)\rho}
{}^{\lambda)}\right)(t\tau_{2})\;.
\end{eqnarray}

Then the tensors $d^{(k)}$ have the form
\begin{eqnarray}
d^{(1)}_{\alpha\beta\mu\nu\gamma\delta\rho\sigma}(t\tau_{1},t\tau_{2})
&=&
-\frac{1}{9}\Big(\Omega_{\beta}{}^{(\iota}\Omega_{\nu}{}^{\kappa)}
M_{\alpha\mu}\Big)(t\tau_{1})\Big(\mathcal{B}_{\delta\sigma}
\Phi_{\gamma\iota}\Phi_{\rho\kappa}\Big)(t\tau_{2})
\nonumber\\
&+&\frac{1}{9}\Big(\mathcal{B}_{\beta\nu}
M_{\alpha\mu}\Big)(t\tau_{1})
\Big(\mathcal{B}_{\delta\sigma}\Omega_{(\gamma\rho)}^{-1}\Big)(t\tau_{2})
\nonumber\\
&+&\frac{1}{9}\Big(\Omega_{\beta}{}^{(\iota}\Omega_{\nu}{}^{\kappa)}
M_{\alpha\mu}\Big)(t\tau_{1})
\Big(\mathcal{A}_{(\delta|\iota|}\mathcal{A}_{\sigma)\kappa}
M_{\gamma\rho}^{-1}\Big)(t\tau_{2})
\nonumber\\
&+&\frac{1}{12}\Big(\Omega_{(\beta}{}^{\iota}
\mathcal{A}_{\nu)\eta}M_{\alpha\mu}\Big)
(t\tau_{1})\Big(
\Omega_{\sigma}{}^{(\eta}\Omega_{\delta}{}^{\epsilon}{}\mathcal{E}_{(1)\rho}{}^{\chi)}
\Phi_{\gamma\iota}\Lambda_{\epsilon\chi}\Big)(t\tau_{2})
\nonumber\\
&-&\frac{1}{24}\Big\{\Big(\mathcal{B}_{\delta\sigma}
M_{\gamma\rho}\Big)(t\tau_{1})
\Big(\Lambda_{\iota\kappa}\Lambda_{\eta\chi}\Big)(t\tau_{2})
\nonumber\\
&+&4\Big(\Omega_{\delta}{}^{(\omega}\Omega_{\sigma}{}^{\lambda)}
M_{\gamma\rho}\Big)(t\tau_{1})
\Big(\Lambda_{\omega\iota}\Lambda_{\lambda\kappa}
\Lambda_{\eta\chi}\Big)(t\tau_{2})\Big\}
\Big(\Omega_{\beta}{}^{(\iota}\Omega_{\nu}{}^{\kappa}
\Psi_{\alpha}{}^{\eta}\mathcal{E}_{(3)\mu}{}^{\chi)}\Big)(t\tau_{2})
\nonumber\\
&+&\frac{1}{3}
\Big(\Omega_{\beta}{}^{(\iota}\Omega_{\nu}{}^{\kappa}\Psi_{\alpha}
{}^{\eta}\mathcal{E}_{(3)\mu}{}^{\chi)}\Big)
(t\tau_{1})\mathcal{V}_{\gamma\delta\rho\sigma\iota\kappa\eta\chi}
(t\tau_{1},t\tau_{2})\;,
\end{eqnarray}
\begin{eqnarray}
d^{(2)}_{\mu\nu\alpha\beta\rho\sigma}(t\tau_{1},t\tau_{2})
&=&
\frac{1}{9}\Big(\Omega_{\beta}{}^{(\iota}\Omega_{\sigma}{}^{\kappa)}
M_{\alpha\rho}\Big)(t\tau_{1})
\Big(\Phi_{\mu\iota}\mathcal{A}_{\nu\kappa}\Big)(t\tau_{2})
\nonumber\\
&-&
\frac{1}{9}\Big(\mathcal{B}_{\beta\sigma}
M_{\alpha\rho}\Big)(t\tau_{1})\mathcal{A}_{(\mu\nu)}(t\tau_{2})
\nonumber\\
&-&\frac{1}{9}\mathcal{A}_{(\mu\nu)}(t\tau_{1})
\Big(\mathcal{B}_{\beta\sigma}M_{\alpha\rho}\Big)(t\tau_{2})
\nonumber\\
&+&
\frac{1}{12}\Big(\Omega_{\beta}{}^{(\iota}
\Omega_{\sigma}{}^{\kappa}\Psi_{\alpha}{}^{\eta}\mathcal{E}_{(3)\rho}{}^{\chi)}
\mathcal{A}_{(\mu\nu)}\Lambda_{\iota\kappa}\Lambda_{\eta\chi}\Big)(t\tau_{2})
\nonumber\\
&-&\frac{1}{36}\Big(\Omega_{\beta}{}^{(\iota}
\Omega_{\sigma}{}^{\kappa)}M_{\alpha\rho}\Big)
(t\tau_{1})\Big(\mathcal{A}_{\nu\iota}\Lambda_{\kappa\eta}
\mathcal{E}_{(7)\mu}{}^{\eta}\Big)(t\tau_{2})
\nonumber\\
&+&\frac{1}{3}\Big(\Omega_{\beta}{}^{(\iota}\Omega_{\sigma}{}^{\kappa}
\Psi_{\alpha}{}^{\eta}\mathcal{E}_{(3)\rho}{}^{\chi)}\Big)(t\tau_{1})
\Bigg\{-\Lambda_{\kappa\eta}(t\tau_{1})\Big(\mathcal{A}_{\nu\chi}
\Phi_{\mu\iota}\Big)(t\tau_{2})
\nonumber\\
&+&\frac{1}{2}\Big(\mathcal{A}_{(\mu\nu)}\Lambda_{\iota\kappa}
\Lambda_{\eta\chi}\Big)(t\tau_{1})
+\frac{1}{4}\Lambda_{\iota\kappa}(t\tau_{1})\Big(
\mathcal{A}_{\nu(\eta}\Lambda_{\chi)\epsilon}\mathcal{E}_{(7)\mu}{}^{\epsilon}\Big)(t\tau_{2})\Bigg\}
\nonumber\\
&+&\frac{1}{36}\Omega_{\nu}{}^{(\iota}\mathcal{E}_{(7)\mu}{}^{\kappa)}
(t\tau_{1})\mathcal{S}_{\alpha\beta\rho\sigma\iota\kappa}(t\tau_{2})\;,
\end{eqnarray}
\begin{eqnarray}
d^{(3)}_{\mu\nu\alpha\beta}(t\tau_{1},t\tau_{2})
&=&
-\frac{1}{36}\Big(\Omega_{\nu}{}^{(\iota}\mathcal{E}_{(7)\mu}
{}^{\kappa)}\Big)(t\tau_{1})\Big(\Phi_{\alpha\iota}
\mathcal{A}_{\beta\kappa}\Big)(t\tau_{2})
+\frac{2}{9}\mathcal{A}_{(\mu\nu)}(t\tau_{1})
\mathcal{A}_{(\alpha\beta)}(t\tau_{2})
\nonumber\\
&+&\frac{1}{144}\Omega_{\nu}{}^{(\iota}\mathcal{E}_{(7)\mu}{}^{\kappa)}
(t\tau_{1})\Big(\mathcal{A}_{\beta\iota}\Lambda_{\kappa\sigma}\mathcal{E}_{(7)\alpha}{}^{\sigma}\Big)(t\tau_{2})\;.
\end{eqnarray}
\begin{eqnarray}
d^{(4)}_{\mu\nu\alpha\beta\rho\sigma}(t\tau_{1},t\tau_{2})
&=&
-\frac{1}{3}\Big(\Omega_{\beta}{}^{(\iota}
\Omega_{\sigma}{}^{\kappa)}M_{\alpha\rho}\Big)
(t\tau_{1})\Big(\Phi_{\mu\iota}\mathcal{A}_{\nu\kappa}\Big)(t\tau_{2})
\nonumber\\
&-&\frac{2}{3}\Big(\Omega_{\nu}{}^{(\iota}
\Psi_{\mu}{}^{\kappa)}\Big)(t\tau_{1})
\Big(\mathcal{B}_{\beta\sigma}\Phi_{\alpha\iota}
\Phi_{\rho\kappa}\Big)(t\tau_{2})
+\frac{1}{3}\Big(\mathcal{B}_{\beta\sigma}
M_{\alpha\rho}\Big)(t\tau_{1})\mathcal{A}_{\mu\nu}(t\tau_{2})
\nonumber\\
&+&\frac{1}{3}\mathcal{A}_{\mu\nu}(t\tau_{1})
\Big(\mathcal{B}_{\beta\sigma}M_{\alpha\rho}\Big)(t\tau_{2})
\nonumber\\
&+&\frac{2}{3}\Big(\Omega_{\beta}{}^{(\iota}
\Omega_{\sigma}{}^{\kappa)}
M_{\alpha\rho}\Big)(t\tau_{1})
\Big(\Psi_{\mu}{}^{\eta}\mathcal{A}_{\nu\iota}\Lambda_{\kappa\eta}
\Big)(t\tau_{2})
\nonumber\\
&+&\frac{2}{3}\Big(\Omega_{\nu}{}^{(\iota}
\Psi_{\mu}{}^{\kappa)}\Big)(t\tau_{1})
\Big(\mathcal{A}_{\beta\iota}\mathcal{A}_{\sigma\kappa}
M_{\alpha\rho}\Big)(t\tau_{2})
\nonumber\\
&+&\frac{1}{2}\Big(\Omega_{\nu}{}^{(\epsilon}
\Psi_{\mu}{}^{\lambda)}\Big)(t\tau_{1})
\Big(\Omega_{\beta}{}^{(\iota}\Omega_{\sigma}
{}^{\kappa}\mathcal{E}_{(1)\rho}{}^{\eta)}\Phi_{\alpha\epsilon}
\Lambda_{\lambda\iota}\Lambda_{\kappa\eta}\Big)(t\tau_{2})
\nonumber\\
&+&\frac{1}{2}\Big(\Omega_{\beta}{}^{(\iota}
\Omega_{\sigma}{}^{\kappa}\Psi_{\alpha}{}^{\eta}\mathcal{E}_{(3)\rho}
{}^{\epsilon)}\Big)(t\tau_{1})\Big\{2\Lambda_{\kappa\eta}(t\tau_{1})
\Big(\mathcal{A}_{\nu\epsilon}\Phi_{\mu\iota}\Big)(t\tau_{2})
\nonumber\\
&-&\Big(\mathcal{A}_{\mu\nu}\Lambda_{\iota\kappa}
\Lambda_{\eta\epsilon}\Big)(t\tau_{1})
-4\Lambda_{\iota\kappa}(t\tau_{1})
\Big(\Psi_{\mu}{}^{\lambda}\mathcal{A}_{\nu\epsilon}\Lambda_{\eta
\lambda}\Big)(t\tau_{2})\Big\}
\nonumber\\
&-&\frac{1}{4}\Big\{\Big(\mathcal{A}_{\mu\nu}
\Lambda_{\iota\kappa}\Lambda_{\eta\epsilon}\Big)(t\tau_{2})
+4\Big(\Omega_{\nu}{}^{(\omega}\Psi_{\mu}{}^{\lambda)}\Big)(t\tau_{1})
\Big(\Lambda_{\omega\iota}\Lambda_{\lambda\kappa}
\Lambda_{\eta\epsilon}\Big)(t\tau_{2})\Big\}
\nonumber\\
&\times&\Big(\Omega_{\beta}{}^{(\iota}\Omega_{\sigma}
{}^{\kappa}\Psi_{\alpha}{}^{\eta}\mathcal{E}_{(3)\rho}{}^{\epsilon)}\Big)(t\tau_{2})\;.
\end{eqnarray}
\begin{eqnarray}
d^{(5)}_{\mu\nu\alpha\beta}(t\tau_{1},t\tau_{2})
&=&
\frac{1}{12}\Big(\Omega_{\beta}{}^{(\gamma}\mathcal{E}_{(7)\alpha}
{}^{\delta)}\Big)(t\tau_{1})\Big(\Phi_{\mu\gamma}
\mathcal{A}_{\nu\delta}\Big)(t\tau_{2})
\nonumber\\
&+&
\frac{2}{3}\Big(\Omega_{\nu}{}^{(\gamma}
\Psi_{\mu}{}^{\delta)}\Big)(t\tau_{1})
\Big(\Phi_{\alpha\gamma}\mathcal{A}_{\beta\delta}\Big)(t\tau_{2})
\nonumber\\
&-&\frac{2}{3}\mathcal{A}_{(\alpha\beta)}(t\tau_{1})
\mathcal{A}_{\mu\nu}(t\tau_{2})
-\frac{2}{3}\mathcal{A}_{(\alpha\beta)}(t\tau_{2})
\mathcal{A}_{\mu\nu}(t\tau_{1})
\nonumber\\
&-&\frac{1}{6}\Big(\Omega_{\beta}{}^{(\iota}
\mathcal{E}_{(7)\alpha}{}^{\kappa)}\Big)(t\tau_{1})\Big(\Psi_{\mu}{}^{\epsilon}
\mathcal{A}_{\nu\kappa}
\Lambda_{\iota\epsilon}\Big)(t\tau_{2})
\nonumber\\
&-&\frac{1}{6}\Big(\Omega_{\nu}{}^{(\iota}
\Psi_{\mu}{}^{\kappa)}\Big)(t\tau_{1})
\Big(\mathcal{C}_{\beta}{}^{(\eta}{}_{\alpha}
{}^{\epsilon)}\Lambda_{\iota\eta}\Lambda_{\kappa\epsilon}\Big)(t\tau_{2})\;.
\end{eqnarray}
\begin{eqnarray}
d^{(6)}_{\mu\nu\alpha\beta}(t\tau_{1},t\tau_{2})&=&
-2\Big(\Omega_{\nu}{}^{(\gamma}\Psi_{\mu}{}^{\delta)}\Big)
(t\tau_{1})\Big(\Phi_{\alpha\gamma}\mathcal{A}_{\beta\delta}\Big)(t\tau_{2})
+2\mathcal{A}_{\mu\nu}(t\tau_{1})\mathcal{A}_{\alpha\beta}(t\tau_{2})
\nonumber\\
&+&4\Big(\Omega_{\nu}{}^{(\gamma}\Psi_{\mu}{}^{\delta)}
\Big)(t\tau_{1})\Big(\Psi_{\alpha}{}^{\sigma}
\mathcal{A}_{\beta\delta}\Lambda_{\gamma\sigma}\Big)(t\tau_{2})\;.
\end{eqnarray}

%========================================================
%=====================================================================

\section{Conclusions}
\setcounter{equation}0

In this paper we studied the heat kernel expansion for a Laplace operator
acting on sections of a complex vector bundle over a smooth compact
Riemannian manifold without boundary. We assumed that the curvature $F$ of
the $U(1)$ part of the total connection (the electromagnetic field) is
covariantly constant and large, so that $tF\sim 1$, that is, $F$ is of order
$t^{-1}$. In this situation the standard asymptotic expansion of the heat
kernel as $t\to 0$ does not apply since the electromagnetic field can not be
treated as a perturbation.

In order to calculate the heat kernel asymptotic expansion we use an algebraic
approach in which the nilpotent algebra of the operators ${\cal D}_\mu$ plays
a major role. In this approach the calculation of the asymptotic expansion of
the heat kernel is reduced to the calculation of the asymptotic expansion of
the heat semigroup and, then, to the action of differential operators on the
zero-order heat kernel. Since the zero-order heat kernel has the Gaussian form
the heat kernel asymptotics are expressed in terms of generalized Hermite
polynomials.

The main result of this work is establishing the existence of a new
non-per\-tur\-bative asymptotic expansion of the heat kernel and the explicit
calculation of the first three coefficients of this expansion (both
off-diagonal and the diagonal ones). As far as we know, such an asymptotic
expansion and the explicit form of these modified heat kernel coefficients are
new.

We presented our result as explicitly as possible. Unfortunately, some of
the integrals of the tensor-valued functions cannot be evaluated explicitly in
full generality. They can be evaluated, in principle, by using the spectral
decomposition of the two-form $F$,
\be
F=\sum_{k=1}^{[n/2]} B_k E_k\,,
\qquad
F^2=-\sum_{k=1}^{[n/2]} B_k^2 \Pi_k\,,
\ee
where $B_k$ are the eigenvalues, $E_k$ are the (2-dimensional) eigen-two-forms,
and $\Pi_k=-E_k^2$ are the corresponding eigen-projections onto 2-dimensional
eigenspaces. Then for any analytic function of $tiF$ we have
\be
f(tiF)=\sum_{k=1}^{[n/2]} f(tB_k) \frac{1}{2}(\Pi_k+iE_k)
+\sum_{k=1}^{[n/2]} f(-tB_k) \frac{1}{2}(\Pi_k-iE_k)\,.
\ee
However, this seems impractical in general case in $n$ dimensions.
It would simplify substantially in the following cases: i) there is only one eigenvalue
(one magnetic field) in a corresponding two-dimensional subspace, that is,
$F=B_1 E_1$ (which is essentially 2-dimensional), and ii) all eigenvalues are
equal so that $F^2=-I$ (which is only possible in even dimensions).
We plan to study this problem in a future work.

The work carried on in this paper can find useful applications in various
fields of theoretical physics and mathematics. For instance, our results
can be applied to the study of the heat kernel asymptotic expansion on
K\"{a}hler manifolds. The complex structure on K\"ahler manifolds is a parallel
antisymmetric two-tensor which plays the role of the covariantly constant
electromagnetic field. This subject is also interesting, in particular, in
connection with String Theory.

%=================================================================
%==============================================
\section*{Appendix A. Hermite Polynomials }
\setcounter{equation}0
\renewcommand\theequation{A.\arabic{equation}}

The Hermite polynomials are defined by
\bea
\mathcal{H}_{\mu_{1}\cdots\mu_{n}}
&=&\exp\left\{-\frac{1}{2}u^\alpha \Lambda_{\alpha\beta}u^\beta\right\}
\frac{\partial}{\partial u^{\mu_1}}\cdots
\frac{\partial}{\partial u^{\mu_n}}
\exp\left\{\frac{1}{2}u^\alpha \Lambda_{\alpha\beta}u^\beta\right\}
\nonumber\\
&=&
\left(\frac{\partial}{\partial u^{\mu_1}}+\Lambda_{\mu_1\nu_1}
u^{\nu_1}\right)\cdots
\left(\frac{\partial}{\partial u^{\mu_n}}+\Lambda_{\mu_n\nu_n}
u^{\nu_n}\right)
\cdot 1\,.
\eea
They can be computed explicitly as follows.
First, let
\begin{equation}\label{59aa}
\mathcal{H}_{(n)}(\xi)
=\xi^{\mu_{1}}\cdots\xi^{\mu_{n}}\mathcal{H}_{\mu_{1}\cdots\mu_{n}}
\end{equation}
and
\begin{equation}\label{59c}
B=\xi^{\mu}\frac{\partial}{\partial u^\mu}\;,\qquad
A=\xi^\mu\Lambda_{\mu\nu}u^\nu\;.
\end{equation}
Then
\begin{equation}\label{60aa}
\mathcal{H}_{(n)}(\xi)
=(A+B)^n\cdot 1\;.
\end{equation}

Finally, let
\be
C=[B,A]=\xi^\mu\Lambda_{\mu\nu}\xi^\nu\,.
\ee
Obviously, the operators $A$, $B$, $C$ form the Heisenberg algebra
\begin{displaymath}
[B,A]=C\;,\quad [A,C]=[B,C]=0\;.
\end{displaymath}

%====================================================
\begin{lemma}\label{lemma1}
There holds,
\begin{equation}\label{60}
(A+B)^{n}=\sum_{k=0}^{\left[\frac{n}{2}\right]}
\sum_{m=0}^{n-2k}\frac{(2k)!}{2^{k}k!}
{n\choose 2k}
\;{n-2k \choose m}
C^{k}A^{n-2k-m}B^{m}\;.
\end{equation}
\end{lemma}
\begin{proof}
Notice that $e^{t(A+B)}$ is the generating functional for $(A+B)^{n}$.
Now, by using the Baker-Hausdorff-Campbell formula
\begin{displaymath}
e^{t(A+B)}=e^{\frac{t^{2}}{2}C}e^{tA}e^{tB}\;,
\end{displaymath}
expanding both sides in $t$ and
computing the Taylor coefficients of the right hand side
we obtain the eq. (\ref{60}).
\end{proof}
%=============================

By using this result we
obtain an explicit expression for (\ref{60aa})
\begin{equation}
\label{61}
\mathcal{H}_{(n)}(\xi)=
\xi^{\mu_{1}}\cdots\xi^{\mu_{n}}\mathcal{H}_{\mu_{1}\cdots\mu_{n}}
=\sum_{k=0}^{\left[\frac{n}{2}\right]}
\frac{n!}{2^{k}k!(n-2k)!}C^{k}A^{n-2k}\;.
\end{equation}
By setting $A=0$ we immediately obtain the (diagonal)
values of Hermite polynomials
at $u=0$
\bea
\left[\mathcal{H}_{\mu_{1}\cdots\mu_{2n+1}}\right]^{\rm diag}&=&0\,,
\\
\left[\mathcal{H}_{\mu_{1}\cdots\mu_{2n}}\right]^{\rm diag}
&=&\frac{(2n)!}{2^n n!}
\Lambda_{(\mu_{1}\mu_{2}}\cdots\Lambda_{\mu_{2n-1}\mu_{2n})}
\,.
\eea

We list below a few low order Hermite polynomials needed for our calculation
\begin{eqnarray}
\mathcal{H}_{(0)}&=&1\;,
\\[10pt]
\mathcal{H}_{\mu_{1}}
&=&\Lambda_{\mu_{1}\alpha}u^{\alpha}\;,
\\[10pt]
\mathcal{H}_{\mu_{1}\mu_{2}}
&=&\Lambda_{(\mu_{1}\mu_{2})}
+\Lambda_{\mu_{1}\alpha}
\Lambda_{\mu_{1}\beta}u^{\alpha}u^{\beta}\;,
\\[10pt]
\mathcal{H}_{\mu_{1}\mu_{2}\mu_{3}}
&=&3\Lambda_{(\mu_{1}\mu_{2}}
\Lambda_{\mu_{3})\alpha}u^{\alpha}
+\Lambda_{\mu_{1}\alpha}
\Lambda_{\mu_{2}\beta}\Lambda_{\mu_{3}\gamma}
u^{\alpha}u^{\beta}u^{\gamma}\;,
\\[10pt]
\mathcal{H}_{\mu_{1}\mu_{2}\mu_{3}\mu_{4}}
&=&3\Lambda_{(\mu_{1}\mu_{2}}\Lambda_{\mu_{3}\mu_{4})}
+3\Lambda_{(\mu_{1}\mu_{2}}
\Lambda_{\mu_{3}|\alpha|}\Lambda_{\mu_{4})\beta}u^{\alpha}u^{\beta}
\nonumber\\
&+&\Lambda_{\mu_{1}\alpha}
\Lambda_{\mu_{2}\beta}\Lambda_{\mu_{3}\gamma}
\Lambda_{\mu_{4}\delta}u^{\alpha}u^{\beta}u^{\gamma}u^{\delta}\;.
%\\[10pt]
\end{eqnarray}
\begin{eqnarray}
\mathcal{H}_{\mu_{1}\mu_{2}\mu_{3}\mu_{4}\mu_{5}}
&=&15\Lambda_{(\mu_{1}\mu_{2}}
\Lambda_{\mu_{3}\mu_{4}}\Lambda_{\mu_{5})\alpha}u^{\alpha}
+5\Lambda_{(\mu_{1}\mu_{2}}
\Lambda_{\mu_{3}|\alpha|}\Lambda_{\mu_{4}|\beta|}
\Lambda_{\mu_{5})\gamma}u^{\alpha}u^{\beta}u^{\gamma}
\nonumber\\
&+&\Lambda_{\mu_{1}\alpha}
\Lambda_{\mu_{2}\beta}\Lambda_{\mu_{3}\gamma}
\Lambda_{\mu_{4}\delta}\Lambda_{\mu_{5}\eta}
u^{\alpha}u^{\beta}u^{\gamma}u^{\delta}u^{\eta}\;,
\\[10pt]
\mathcal{H}_{\mu_{1}\mu_{2}\mu_{3}\mu_{4}\mu_{5}\mu_{6}}
&=&15\Lambda_{(\mu_{1}\mu_{2}}
\Lambda_{\mu_{3}\mu_{4}}\Lambda_{\mu_{5}\mu_{6})}
+45\Lambda_{(\mu_{1}\mu_{2}}
\Lambda_{\mu_{3}\mu_{4}}\Lambda_{\mu_{5}|\alpha|}
\Lambda_{\mu_{6})\beta}u^{\alpha}u^{\beta}
\nonumber\\
&+&
15\Lambda_{(\mu_{1}\mu_{2}}
\Lambda_{\mu_{3}|\alpha|}\Lambda_{\mu_{4}|\beta|}
\Lambda_{\mu_{5}|\gamma|}\Lambda_{\mu_{6})\delta}
u^{\alpha}u^{\beta}u^{\gamma}u^{\delta}
\nonumber\\
&+&
\Lambda_{(\mu_{1}|\alpha|}
\Lambda_{\mu_{2}|\beta|}\Lambda_{\mu_{3}|\gamma|}
\Lambda_{\mu_{4}|\delta|}\Lambda_{\mu_{5}|\eta|}
\Lambda_{\mu_{6})\iota}u^{\alpha}u^{\beta}u^{\gamma}
u^{\delta}u^{\eta}u^{\iota}\;,
\end{eqnarray}

%===========================================
We list below some of the generalized Hermite polynomials of second kind.
Now we have two sets of Hermite polynomials that depend on the
quadratic forms $\Lambda$ at two different times, $s_1$ and
$s_2$.
Let us define
\begin{equation}
\mathcal{H}_{(n)}(s_{1})=
\xi^{\mu_{1}}\cdots\xi^{\mu_{n}}\mathcal{H}_{\mu_{1}\cdots\mu_{n}}(s_{1})\;,
\end{equation}
\begin{equation}
\mathcal{H}_{(n)}(s_{2})=
\eta^{\mu_{1}}\cdots\eta^{\mu_{n}}\mathcal{H}_{\mu_{1}\cdots\mu_{n}}(s_{2})\;,
\end{equation}
and
\begin{equation}
\Lambda(s_{2})=\xi^{\alpha}\Lambda_{\alpha\beta}(s_{2})\eta^{\beta}\;.
\end{equation}
Then from eq. (\ref{613iga}) we obtain
the quantities $\Xi_{(m,n)}$ that we need in our calculations
\begin{eqnarray}
\Xi_{(0,1)}(s_{1},s_{2})
&=&\mathcal{H}_{(1)}(s_{2})
\\
\Xi_{(1,1)}(s_{1},s_{2})
&=&\Lambda(s_{2})
+\mathcal{H}_{(1)}(s_{1})\mathcal{H}_{(1)}(s_{2})\;,
\\
\Xi_{(2,1)}(s_{1},s_{2})
&=&2\Lambda(s_{2})\mathcal{H}_{(1)}(s_{1})
+\mathcal{H}_{(1)}(s_{2})\mathcal{H}_{(2)}(s_{1})\;,
\\
\Xi_{(3,1)}(s_{1},s_{2})
&=&3\Lambda(s_{2})\mathcal{H}_{(2)}(s_{1})
+\mathcal{H}_{(1)}(s_{2})\mathcal{H}_{(3)}(s_{1})\;,
\\
\Xi_{(4,1)}(s_{1},s_{2})
&=&4\Lambda(s_{2})\mathcal{H}_{(3)}(s_{1})
+\mathcal{H}_{(1)}(s_{2})\mathcal{H}_{(4)}(s_{1})\;.
\end{eqnarray}
\begin{eqnarray}
\Xi_{(0,2)}(s_{1},s_{2})
&=&\mathcal{H}_{(2)}(s_{2})
\\
\Xi_{(1,2)}(s_{1},s_{2})
&=&2\Lambda(s_{2})\mathcal{H}_{(1)}(s_{2})
+\mathcal{H}_{(2)}(s_{2})\mathcal{H}_{(1)}(s_{1})\;,
\\
\Xi_{(2,2)}(s_{1},s_{2})
&=&2\Lambda^{2}(s_{2})
+4\Lambda(s_{2})\mathcal{H}_{(1)}(s_{2})\mathcal{H}_{(1)}(s_{1})
+\mathcal{H}_{(2)}(s_{2})\mathcal{H}_{(2)}(s_{1})\;,
\\
\Xi_{(3,2)}(s_{1},s_{2})
&=&6\Lambda^{2}(s_{2})\mathcal{H}_{(1)}(s_{1})
+6\Lambda(s_{2})\mathcal{H}_{(1)}(s_{2})\mathcal{H}_{(2)}(s_{1})
\nonumber\\
&+&\mathcal{H}_{(2)}(s_{2})\mathcal{H}_{(3)}(s_{1})\;,
\\
\Xi_{(4,2)}(s_{1},s_{2})
&=&12\Lambda^{2}(s_{2})\mathcal{H}_{(2)}(s_{1})
+8\Lambda(s_{2})\mathcal{H}_{(1)}(s_{2})\mathcal{H}_{(3)}(s_{1})
\nonumber\\
&+&\mathcal{H}_{(2)}(s_{2})\mathcal{H}_{(4)}(s_{1})\;.
\end{eqnarray}
\begin{eqnarray}
\Xi_{(0,3)}(s_{1},s_{2})
&=&\mathcal{H}_{(3)}(s_{2})
\\
\Xi_{(1,3)}(s_{1},s_{2})
&=&3\Lambda(s_{2})\mathcal{H}_{(2)}(s_{2})
+\mathcal{H}_{(3)}(s_{2})\mathcal{H}_{(1)}(s_{1})\;,
\\
\Xi_{(2,3)}(s_{1},s_{2})
&=&6\Lambda^{2}(s_{2})\mathcal{H}_{(1)}(s_{2})
+6\Lambda(s_{2})\mathcal{H}_{(2)}(s_{2})\mathcal{H}_{(1)}(s_{1})
\nonumber\\
&+&\mathcal{H}_{(3)}(s_{2})\mathcal{H}_{(2)}(s_{1})\;,
\\
\Xi_{(3,3)}(s_{1},s_{2})
&=&6\Lambda^{3}(s_{2})
+18\Lambda^{2}(s_{2})\mathcal{H}_{(1)}(s_{2})\mathcal{H}_{(1)}(s_{1})
\nonumber\\
&+&
9\Lambda(s_{2})\mathcal{H}_{(2)}(s_{2})\mathcal{H}_{(2)}(s_{1})
+\mathcal{H}_{(3)}(s_{2})\mathcal{H}_{(3)}(s_{1})\;,
\\
\Xi_{(4,3)}(s_{1},s_{2})
&=&24\Lambda^{3}(s_{2})\mathcal{H}_{(1)}(s_{1})
+36\Lambda^{2}(s_{2})\mathcal{H}_{(1)}(s_{2})\mathcal{H}_{(2)}(s_{1})
\nonumber\\
&+&
12\Lambda(s_{2})\mathcal{H}_{(2)}(s_{2})\mathcal{H}_{(3)}(s_{1})
+\mathcal{H}_{(3)}(s_{2})\mathcal{H}_{(4)}(s_{1})\;,
\end{eqnarray}
\begin{eqnarray}
\Xi_{(0,4)}(s_{1},s_{2})
&=&\mathcal{H}_{(4)}(s_{2})
\\
\Xi_{(1,4)}(s_{1},s_{2})
&=&4\Lambda(s_{2})\mathcal{H}_{(3)}(s_{2})
+\mathcal{H}_{(4)}(s_{2})\mathcal{H}_{(1)}(s_{1})\;,
\\
\Xi_{(2,4)}(s_{1},s_{2})
&=&12\Lambda^{2}(s_{2})\mathcal{H}_{(2)}(s_{2})
+8\Lambda(s_{2})\mathcal{H}_{(3)}(s_{2})\mathcal{H}_{(1)}(s_{1})
\nonumber\\
&+&\mathcal{H}_{(4)}(s_{2})\mathcal{H}_{(2)}(s_{1})\;,
\\
\Xi_{(3,4)}(s_{1},s_{2})
&=&24\Lambda^{3}(s_{2})\mathcal{H}_{(1)}(s_{2})
+36\Lambda^{2}(s_{2})\mathcal{H}_{(2)}(s_{2})\mathcal{H}_{(1)}(s_{1})
\nonumber\\
&+&
12\Lambda(s_{2})\mathcal{H}_{(3)}(s_{2})\mathcal{H}_{(2)}(s_{1})
+\mathcal{H}_{(4)}(s_{2})\mathcal{H}_{(3)}(s_{1})\;,
\\
\Xi_{(4,4)}(s_{1},s_{2})
&=&24\Lambda^{4}(s_{2})
+96\Lambda^{3}(s_{2})\mathcal{H}_{(1)}(s_{2})\mathcal{H}_{(1)}(s_{1})
+72\Lambda^{2}(s_{2})\mathcal{H}_{(2)}(s_{2})\mathcal{H}_{(2)}(s_{1})
\nonumber\\
&+&16\Lambda(s_{2})\mathcal{H}_{(3)}(s_{2})\mathcal{H}_{(3)}(s_{1})
+\mathcal{H}_{(4)}(s_{2})\mathcal{H}_{(4)}(s_{1})\;,
\end{eqnarray}

The coincidence limit of the quantities $\Xi_{(m,n)}$ ,with $m+n$ odd, vanishes identically
\be
\left[\Xi_{(m,n)}(s_{1},s_{2})\right]^{\rm diag}=0\,,
\qquad \mbox{if $(m+n)$ is odd}
\ee
By recalling the coincidence limits of the Hermite polynomials we obtain the following
\begin{eqnarray}
\left[\Xi_{(1,1)}(s_{1},s_{2})\right]^{\rm diag}
&=&\Lambda(s_{2})\;,
\\
\left[\Xi_{(3,1)}(s_{1},s_{2})\right]^{\rm diag}
&=&3\Lambda(s_{1})\Lambda(s_{2})\;,
\\
\left[\Xi_{(0,2)}(s_{1},s_{2})\right]^{\rm diag}
&=&\Lambda(s_{2})\;,
\\
\left[\Xi_{(2,2)}(s_{1},s_{2})\right]^{\rm diag}
&=&\Lambda(s_{1})\Lambda(s_{2})
+2\Lambda^{2}(s_{2})\;,
\\
\left[\Xi_{(4,2)}(s_{1},s_{2})\right]^{\rm diag}
&=&3\Lambda^{2}(s_{1})\Lambda(s_{2})
+12\Lambda(s_{1})\Lambda^{2}(s_{2})\;,
\\
\left[\Xi_{(1,3)}(s_{1},s_{2})\right]^{\rm diag}
&=&3\Lambda^{2}(s_{2})\;,
\\
\left[\Xi_{(3,3)}(s_{1},s_{2})\right]^{\rm diag}
&=&9\Lambda(s_{1})\Lambda^{2}(s_{2})
+6\Lambda^{3}(s_{2})\;,
\\
\left[\Xi_{(2,4)}(s_{1},s_{2})\right]^{\rm diag}
&=&3\Lambda(s_{1})\Lambda^{2}(s_{2})
+12\Lambda^{3}(s_{2})\;,
\\
\left[\Xi_{(4,4)}(s_{1},s_{2})\right]^{\rm diag}
&=&9\Lambda^{2}(s_{1})\Lambda^{2}(s_{2})
+72\Lambda(s_{1})\Lambda^{3}(s_{2})
+24\Lambda^{4}(s_{2})\;.
\end{eqnarray}

%======================================================
\section*{Appendix B. Commutators}
\setcounter{equation}0
\renewcommand\theequation{B.\arabic{equation}}

\begin{lemma}
Let $D_{\mu}$ and $u^\nu$ be operators satisfying the algebra
\be
[D_\mu, u^\nu]=\delta^\nu_\mu\,,\qquad
[D_\mu,D_\nu]=[u^\mu,u^\nu]=0\,.
\label{800}
\ee
Then
\begin{eqnarray}
\left[D_{\mu_{1}}\cdots D_{\mu_{n}},u^{\rho}\right]
&=&n\;\delta^{\rho}{}_{(\mu_{1}}D_{\mu_{2}}
\cdots D_{\mu_{n})}\;
\label{36b}
\\
\left[D_{\mu_{1}}\cdots D_{\mu_{n}},u^{\rho}u^{\sigma}\right]&=&n(n-1)
\delta^{\rho}{}_{(\mu_{1}}
\delta^{\sigma}{}_{\mu_{2}}D_{\mu_{3}}\cdots D_{\mu_{n})}
\label{36c}
\nonumber\\
&+&2n\;u^{(\rho}\delta^{\sigma)}{}_{(\mu_{1}}D_{\mu_{2}}
\cdots D_{\mu_{n})}\;.
\end{eqnarray}
\end{lemma}

\begin{proof}
Let $\mathcal{X}(\xi)=\xi^{\mu}D_{\mu}$ and
\begin{equation}
\varphi^{\rho}(t)
=\left[e^{t\mathcal{X}(\xi)},u^{\rho}\right]
=\left(e^{t\mathcal{X}(\xi)}u^{\rho}
e^{-t\mathcal{X}(\xi)}
-u^{\rho}\right)e^{t\mathcal{X}(\xi)}\;,
\end{equation}
Then
\begin{equation}
e^{t\mathcal{X}(\xi)}u^{\rho}e^{-t\mathcal{X}(\xi)}
=\sum_{k=0}^{\infty}\frac{t^{k}}{k!}
\left(\textrm{Ad}_{\mathcal{X}(\xi)}\right)^{k}u^{\rho}\;.
\end{equation}
By using the commutation relation in (\ref{800}) we have
\be
[\mathcal{X}(\xi),u^\rho]=\xi^\rho
\ee
and, therefore,
\begin{equation}
e^{t\mathcal{X}(\xi)}u^{\rho}e^{-t\mathcal{X}(\xi)}=u^{\rho}+t\xi^{\rho}\;.
\end{equation}
Thus
\begin{equation}
\varphi^{\rho}(t)=t\xi^{\rho}e^{t\mathcal{X}(\xi)}\;.
\end{equation}
By expanding in Taylor series both sides of the last equation we obtain
\be
\sum_{k=0}^{\infty}
\frac{t^{k+1}}{(k+1)!}\xi^{\mu_{1}}\cdots\xi^{\mu_{k+1}}
\left[D_{(\mu_{1}}\cdots D_{\mu_{k+1})},u^{\rho}\right]
%\nonumber\\
%&=&
=\sum_{k=0}^{\infty}\frac{t^{k+1}}{k!}\xi^{\mu_{1}}
\cdots\xi^{\mu_{k+1}}\delta^{\rho}{}_{(\mu_{1}}D_{\mu_{2}}
\cdots D_{\mu_{k+1})}\;.
\ee
Now by equating the same powers of $t$ in both series we obtain the
claim (\ref{36b}).

The second relation can be proved in a similar manner. We introduce, in this
case, the following generating function
\begin{equation}
\varphi^{\rho\sigma}(t)=\left[e^{t\mathcal{X}(\xi)},u^{\rho}u^{\sigma}\right]
\;.
\end{equation}
By the same argument used in the proof of the first relation we obtain that
\begin{equation}
\varphi^{\rho\sigma}(t)
=\left[e^{t\mathcal{X}(\xi)},u^{\rho}u^{\sigma}\right]
=2t\xi^{(\rho}u^{\sigma)}e^{t\mathcal{X}(\xi)}
+t^{2}\xi^{\rho}\xi^{\sigma}\;.
\end{equation}
Now, as before, by expanding the last equation in Taylor series and equating
the same powers of $t$
we obtain the claim (\ref{36c}).
\end{proof}

%==========================================================

\end{document}